\documentclass[journal=jtcce,manuscript=article]{achemso}

\usepackage{bm}
\usepackage{chemformula} 
\usepackage[T1]{fontenc} 
\usepackage{booktabs}
\usepackage{multirow}

\author{Kaihan Lin}

\author{Xing Gao}
\affiliation
{ School of Materials, Sun Yat-sen University, Shenzhen, Guangdong 518107, China}
\email{gxing@mail.sysu.edu.cn}

\title[An \textsf{achemso} demo]
  {Enhancing Open Quantum Dynamics Simulations Using Neural Network-Based Non-Markovian Stochastic Schrödinger Equation Method}

\abbreviations{IR,NMR,UV}

\begin{document}


\begin{abstract}
The Non-Markovian Stochastic Schrödinger Equation (NMSSE) offers a promising approach for open quantum simulations, especially in large systems, owing to its low scaling complexity and suitability for parallel computing. However, its application at low temperatures faces significant convergence challenges. While short-time evolution converges quickly, long-time evolution requires a much larger number of stochastic trajectories, leading to high computational costs.
To this end, we propose a scheme that combines neural network techniques with simulations of the non-Markovian stochastic Schrödinger equation.
By integrating convolutional neural networks (CNNs) and long short-term memory recurrent neural networks (LSTMs), along with the iterative attentional feature fusion (iAFF) technique, this approach significantly reduces the number of trajectories required for long-time simulations-particularly at low temperatures—thereby substantially lowering computational costs and improving convergence.
To demonstrate our approach, we investigated the dynamics of the spin-boson model and the Fenna-Matthews-Olson (FMO) complex across a range of parameter variations.
\end{abstract}

\section{\label{sec:level1}Introduction}

Open quantum dynamics, achieved by averaging the bath degrees of freedom to derive equations for the system of interest, is widely used to study various physical and chemical systems\cite{ct1,ct2,s3,s4,s5,openquan1,openquan2}.
Accurate simulation of these open quantum systems requires accounting for environmental memory effects, also known as non-Markovian effects.\cite{Grabert:1988yt,PhysRevA.50.3650,PhysRevD.51.1577}
Various non-Markovian dynamic methods based on reduced density matrix, such as the hierarchical equation of motion (HEOM)\cite{heom,heom2,heom3} , the quasi-adiabatic path integral (QUAPI)\cite{quapi1,quapi2} , and the Nakajima-Zwanzig equation\cite{nk1,nk2}, have been proposed. Another alternative description for open quantum dynamics, known as the Stochastic Schrödinger Equation (SSE)\cite{sse1,sse2,sse3}, is based on wavefunction. Ensemble averages across stochastic trajectories generated by SSE reproduce the dynamics of the original reduced density matrix. 
The advantage of the SSE lies in its linear scaling of variable numbers with the number of basis states, unlike the quadratic scaling seen in density matrix equations, which is particularly promising for large systems. 

In particular, Diósi and Strunz proposed the non-Markovian quantum state diffustion (NMQSD) method\cite{nmsse1} to account for non-Markovian effects in SSE. 
However, solving the functional derivative term in NMQSD is challenging. To address this issue, auxiliary stochastic wave functions were introduced to construct hierarchical stochastic equations, termed the hierarchy of pure states (HOPS) method\cite{hops}. 
To reduce the number of auxiliary wave functions in HOPS and improve the feasibility, new noise schemes and different expansions of bath correlation functions have significantly improved.\cite{temperature,expansion1,pade,noise,noise2} 
Additionally, the introduction of pseudo-Fock states and state scaling technique led to the derivation of a non-Markovian stochastic Schrödinger equation with complex frequency modes (cNMSSE)\cite{cnmsse}. 
The sum of products forms of the operators in cNMSSE make it easy to implement with the matrix product state (MPS) technique\cite{mps,cnmsse}.
These developments have enabled a more effective handling of complex quantum processes, such as exciton-phonon interactions in excitation energy transfer, carrier transport in organic semiconductors, and the spectra or charge transport properties of organic materials.\cite{ex-ph,ca-tran1,ca-tran2,spectra,cnmsse}

The stochastic quantum state diffusion method encounters convergence challenges, especially in low-temperature regimes. While the ensemble average converges quickly during short-term evolution, a significantly larger ensemble is needed to achieve accurate convergence over longer time scales.
This numerical instability at long-term simulations hinders our ability to obtain a non-equilibrium steady state, which is crucial for understanding numerous exotic structures, such as nontrivial topology\cite{ss-ntopo1,ss-ntopo2}, and novel dissipative phases of matter\cite{ss-f1,ss-f2,ss-f3,ss-f4}.

A promising approach to address the convergence issue involves utilizing the short-term simulation results from the stochastic method to construct nonlinear dynamical maps for processing long-term evolution, thus significantly reducing the number of required stochastic trajectories. Previous studies on the transfer tensor method (TTM)\cite{ttm} and the memory kernel superoperator within the generalized quantum master equation (GQME) framework\cite{mk} have verified that essential information for future propagation is contained in short time series. By extracting correlation information from these initial stages, nonlinear dynamical maps can be constructed to process subsequent evolution. However these dynamical maps must be acquired through an external exact numerical method. As the system size increases, the computational cost of generating these maps rises significantly. Breakthroughs in neural network technology offer a cost-effective alternative. 

For time-series problems, neural networks that store previous historical information through recurrent connections, such as Recurrent Neural Networks (RNNs)\cite{rnn}, are particularly suitable for training as the nonlinear dynamical maps. However, vanishing and exploding gradient problems are encountered in long-time scale situations\cite{rnnlimit}. To overcome this limitation, a variation of RNN-based model, Long Short-Term Memory Recurrent Neural Networks (LSTMs)\cite{lstm}, can preserve long-range dependencies in sequential data. To extract spatial patterns from multivariate time-series data, incorporating convolutional neural networks (CNNs)\cite{cnn0} with LSTMs is a widely used approach. Additionally, Gated Recurrent Unit (GRU)\cite{gru}, Temporal Convolutional Network (TCN)\cite{tcn}, Transformer\cite{trans}, CNNs, Kernel Ridge Regression (KRR)\cite{krr}, and any other combinations of these neural networks trained as non-linear maps can serve similar roles.

In recent years, time-series forecasting neural networks have been utilized to accelerate various open quantum dynamic algorithms. By using GRUs for defining Lindblad operators, a method within the framework of the Gorini–Kossakowski–Sudarshan–Lindblad master equation has been proposed.\cite{rnn-open} A multi-optimized RNNs, trained by data from the master equation, have been utilized to predict the dynamics of photosynthetic excitation energy transfer in a light-harvesting complex.\cite{rnn-exc}. Thermal correction to wave-packet dynamics has also been learned by deep neural networks to improve the numerical integration of the Schrödinger equation.\cite{ml-correction} Additionally, non-linear autoregressive neural networks have been adopted to support the simulations of Dirac-Frenkel time-dependent variation with the multiple Davydov D2 Ansatz\cite{ml+open2}. KRR has been adopted to accelerate HEOM method for long-term dynamics.\cite{krr} CNNs have been used to assist local thermalising Lindblad master equation and HEOM to propagate the quantum dissipation trajectories.\cite{cnn-exc,cnn} The extension of standard LSTMs, Bidirectional LSTMs (BLSTMs) and combination with CNNs (CLSTMs), are applied in linearized semiclassical mapping dynamics and symmetrical quasi-classical mapping dynamics\cite{ml+open4}. These machine learning applications demonstrate high efficiency, significantly accelerating traditional open quantum dynamics algorithms. \cite{rbm1,rbm2,rbm3,rbm4,ml+open1,ml+open3,cnn-oneshot}

\begin{figure*}
\includegraphics[scale=0.5]{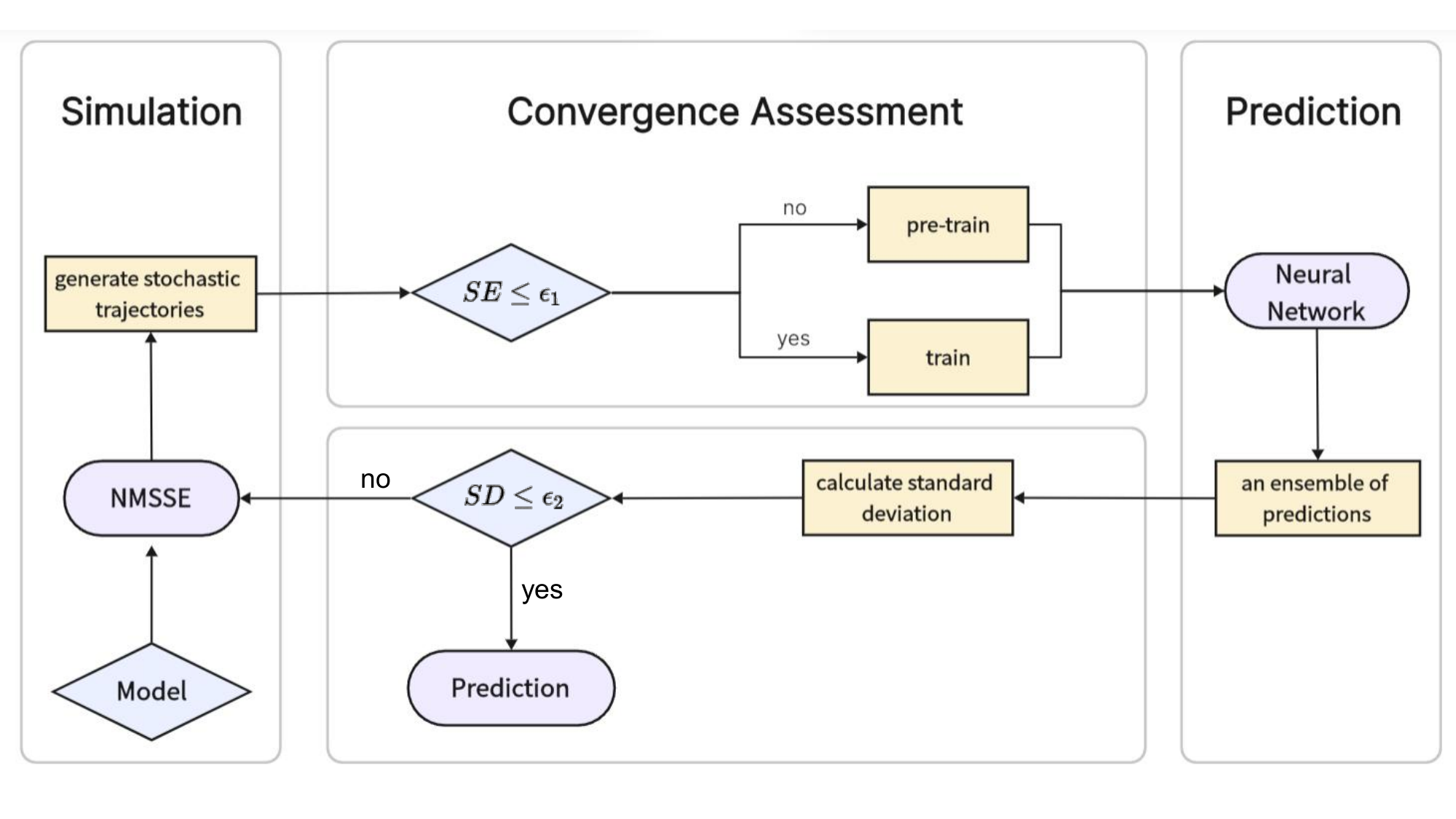}
\caption{\label{fig:sche}The diagram demonstrates the enhancement of open quantum dynamic simulations using neural networks within the NMSSE framework. In the figure, SE represents the standard error calculated across the stochastic trajectories, while SD indicates the standard deviation measured over the ensemble of neural network predictions.}
\end{figure*}


In this work, we integrate neural network techniques with the NMSSE method to address convergence challenges. However, several key issues arise. Determining the appropriate length of time-series data from NMSSE to use as training input for the neural network remains uncertain. Additionally, without access to future data, it is challenging to assess the accuracy of the neural network's predictions. To overcome these problems, we propose the Neural Network-based Non-Markovian Stochastic Schrödinger Equation (NN-NMSSE) method, offering a comprehensive algorithmic solution. The NN-NMSSE framework is composed of 4 key components: generating raw trajectories through NMSSE, evaluating trajectory convergence, training and selecting neural networks, and analyzing the stability of predicted results (as shown in Figure~\ref{fig:sche}). 
This approach goes beyond merely integrating neural networks into the NMSSE method. Instead, it involves an iterative process to progressively refine and stabilize the predictions. The detailed scheme can be seen in \ref{sec:scheme}Methodology. 
cNMSSEs simulations using matrix product states are performed on the Renormalizer package \cite{renormalizer}. Neural networks implemented in TensorFlow\cite{tensorflow} will be used to extract dynamic correlation features and manage long-term evolutions.
The validity of our proposed scheme will be verified through simulations of benchmark models, including the spin-Boson model (SBM)\cite{sbm} and the excitonic dynamics in the Fenna-Matthews-Olson (FMO)\cite{fmo1,fmo} complex, a well-studied light-harvesting system in green sulfur bacteria. Numerical results demonstrate our approach markedly reduces the requisite number of stochastic trajectories.

In the following sections, we will first illustrate the NMSSE method used to collect raw data of the quantum dynamics and then demonstrate how to apply deep learning technique to and address the stochastic oscillations encountered in open quantum system simulations at low temperatures. Detailed settings of the neural networks will be attached in the appendix.

\section{\label{sec:leve2}methodology}

\subsection{\label{sec:leve21}1.Non-Markovian Stochastic Schrödinger Equations with complex modes(cNMSSEs)}

For an open quantum system, the total Hamiltonian can be written as
\begin{equation}
    \hat{H}=\hat{H}_{S}+\hat{H}_{B}+\hat{H}_{SB},
\end{equation}
where $\hat{H}_{S}=\frac{\hat{p}^{2}}{2m_{s}}+V_{s}(\hat{q}),$ $\hat{H}_{B}=\sum_{j=1}^{N}\left(\frac{\hat{p}_{j}^{2}}{2}+\frac{1}{2}\omega_{j}^{2}\hat{x}_{j}^{2}\right), $$\hat{H}_{SB}=f(\hat{q})\otimes\hat{F}=f(\hat{q})\otimes\sum_{j=1}^{N}c_{j}\hat{x}_{j}. $ 
$\hat q$ and $\hat p$ represent the coordinate and momentum of the system's degree of freedom (DOF). Additionally, $\hat x_j$ and $\hat p_j$ denote the coordinate and momentum of the $j$th bath with frequency $\omega_j$. The strength of the coupling between the system and the $j^{\text{th}}$ bath mode is represented by $c_j$, with the discrete spectral density:
\begin{equation}
    J(\omega)=\frac{\pi}{2}\sum_{j=1}^{N}\frac{c_{j}^{2}}{\omega_{j}}\delta(\omega-\omega_{j}).
\end{equation}

To simulate the quantum dynamics, we employed the forward and backward Non-Markovian Stochastic Schrödinger Equations with complex modes\cite{cnmsse} (cNMSSEs):
\begin{align}
    \frac{d}{dt}|\phi^f(t)\rangle=-i\hat{H}^f_{\mathrm{eff}}|\phi^f(t)\rangle\\
    \frac{d}{dt}|\phi^b(t)\rangle=-i\hat{H}^b_{\mathrm{eff}}|\phi^b(t)\rangle
\label{eq:cnmsse}
\end{align}
The quantum state $|\phi^{(f,b)}(t)\rangle$ consists of the vectors of HOPS $\Psi_\mathbf{n}$ and pseudo-Fock states $|\mathbf{n}\rangle$:
\begin{equation}
    |\phi(t)\rangle=\sum_n|\Psi^n(\xi,t)\rangle\otimes|n\rangle 
\end{equation}
where $|\mathbf{n}\rangle=|n_{1},\cdots,n_{K}\rangle$ with the orthonormal relation $\langle\mathbf{n}|\mathbf{n}^{\prime}\rangle=\delta_{\mathbf{nn}^{\prime}}$ and the effective Hamiltonian $\hat{H}^{(f,b)}_{eff}$ can be written as:
\begin{align}   \hat{H}^f_{\mathrm{eff}}=\hat{H}_{s}+\hat{f}(q)\xi_1(t)-i\sum^K_i v_k \hat{n}_k-\hat{f}(q)\sum^K_i {\sqrt{2d_k}}\hat{p}_k\\
\hat{H}^b_{\mathrm{eff}}=\hat{H}_{s}+\hat{f}(q)\xi_2(t)-i\sum^K_i v_k\hat{n}_k-\hat{f}(q)\sum^K_i {\sqrt{2d_k}}\hat{p}_k
\label{eq:heff}
\end{align}
where the occupation number operator $\hat n_k = \hat{b}_k^{\dagger}\hat{b}_k$, and the momentum operator $\hat{p}_k=\frac{i}{\sqrt{2}}(\hat{b}_k^{\dagger}-\hat{b}_k)$. The creation $\hat b_k^{\dagger}$ and annihilation $\hat b_k$ are introduced as:
 
\begin{equation}
    \begin{aligned}
&\hat{b}_k^{\dagger}|\mathbf{n}\rangle= \sqrt{n_k+1}\left|\mathbf{n}_k^+\right\rangle   \\
&\hat{b}_k\left|\mathbf{n}\right\rangle= \sqrt{n_k}\left|\mathbf{n}_k^-\right\rangle. 
\end{aligned}
\end{equation}
where $\left|\mathbf{n}_k^{\pm}\right\rangle=\left|\mathbf{n}\right\rangle\pm|\mathbf{e}_k\rangle$,$|\mathbf{e}_k\rangle=|0,0,\cdots,1_k,\cdots\rangle$.

In the effective Hamiltonian, $v_k$ and $d_k$ are obtained from the exponential expansion:
\begin{equation}
\tilde{\alpha}(t-s)=\sum_{k=0}^{\infty}d_{k}e^{-\nu_{k}(t-s)}.
\label{eq:expansion}
\end{equation}
where $\tilde{\alpha}(t-s)=\alpha(t-s)-\alpha_{1}(t-s)$. In this formula, $\alpha(t)$ is the bath correlation function: $\alpha(t)=\int_{0}^{\infty}\mathrm{d}\omega\frac{J(\omega)}{\pi}[\coth(\frac{\beta\omega}{2})\cos(\omega t)-i\sin(\omega t)]$. And $\alpha_1(t-s)$ is the statistical property of random noise $\xi_1(t)$ and $\xi_2(t)$:
\begin{align}
\langle \xi_1(t)\rangle=\langle \xi^*_2(t)\rangle=0, \\
\langle \xi_1(t)\xi_1(s)\rangle=\alpha_{1}(t-s),\\
\langle \xi_2^*(t)\xi_2^*(s)\rangle=\alpha_{1}(t-s), \\
\langle \xi_1(t)\xi_2^{*}(s)\rangle=\alpha_{2}(t-s), \\
\langle \xi_2^{*}(t)\xi_1(s)\rangle=\alpha^*_{2}(t-s),
\end{align}

In the subsequent numerical calculations, the Gaussian stochastic processes proposed by Zhao and coworkers\cite{noise} are applied: $\xi_{1}(t)=\chi_{c}(t)+\chi_{1}(t), \xi_{2}^{*}(t)=\chi_{c}^{*}(t)+\chi_{2}(t)$, which are constructed by independent Gaussian white noises $\mu_k^1,\ldots,\mu_k^6$ with properties $\langle\mu_{k}^{i}\rangle=0$ and $\langle\mu_{k}^{i}\mu_{k^{\prime}}^{j}\rangle=\delta_{ij}\delta(k-k^{\prime})$:
\begin{equation}
\begin{aligned}
\chi_{c}(t)& =\int_{0}^{k_{max}}\mathrm{d}kh(k)\left[\sqrt{(g_{k}+1)/2}\left(\mu_{k}^{1}+i\mu_{k}^{2}\right)e^{i\omega_{k}t}\right.\left.+\sqrt{g_{k}/2}\left(\mu_{k}^{1}-i\mu_{k}^{2}\right)e^{-i\omega_{k}t}\right],\\
\chi_{1}(t)& =\int_0^{k_{max}}\mathrm{d}kh(k)\left(\sqrt{g_k+1}-\sqrt{g_k}\right)\times\left(\mu_k^3\cos(\omega_kt)+\mu_k^4\sin(\omega_kt)\right),\\
\chi_2(t)&=\int_0^{k_{max}}\mathrm{d}kh(k)\left(\sqrt{g_k+1}-\sqrt{g_k}\right)\times\left(\mu_k^5\cos(\omega_kt)+\mu_k^6\sin(\omega_kt)\right).
\end{aligned}
\end{equation}
where $g_k=1/(e^{\beta\omega_k}-1)$ is the Gaussian distribution function. Due to the statistical properties
$\langle\xi_1(t)\xi_1(s)\rangle=\langle\xi_2^*(t)\xi_2^*(s)\rangle=\alpha_r(t-s)$, only the imaginary part of the bath correlation function is preserved, resulting in only one exponential decaying term remaining.

Further details for the derivation are provided in the appendix.

\subsection{2.Deep Learning Technique}
\subsubsection{Convolution Neural Networks (CNNs)}

CNNs are effective at extracting localized features from data through convolutional kernels. In open quantum dynamics evolution, the time-dependent correlations of the impact of environment and the system dynamics are inherently localized, especially in non-Markovian processes. 
To capture these localized memory effects, CNNs can be effectively applied.

In the one-dimensional convolutional layer, each output neuron $y_j$ from output vector $\boldsymbol{y}=(y_1,y_2,\cdots,y_Q)$ connects to only a subset of the input neurons $\boldsymbol{x}=(x_1,x_2,\cdots,x_P)$ , allowing the convolutional layer to capture local features. As described in Eq.(\ref{eq:neural connect}), the input neuron $\{x_i\}$ and output neuron $\{y_j\}$ are linked by weights $\{k_i\}$ and biases $\{b_j\}$. The size $p$ of the convolution kernel determines how many input neurons are linked to each output neuron. And the convolution kernel slides by the stride $r$. An one-dimensional convolution kernel of size $p$ connects $p$ input neurons to a single output neuron:
\begin{equation}
  y_{j} = \sum_{i=1+(j-1)\cdot r}^{p+(j-1)\cdot r} k_{i} x_{i} + b_{j}
  \label{eq:neural connect}
\end{equation}

The convolutional structure excels at capturing local features because it maps local neurons to specific output neurons.

In our scheme, we construct a convolutional neural network framework with two layers. The first layer is a downsampling layer that divides the input neurons into ten segments, with each segment connected to one output neuron.  The input vector has a length of $L$, and the convolutional kernel size is set to $0.1 \times(L-1)$ with a stride of $0.1\times L$.

\subsubsection{Long Short-Term Memory Recurrent Neural Networks (LSTMs)}

LSTMs are particularly well-suited for time series forecasting due to their inherent structure, which retains long-term dependencies in the data. Given these strengths, we employ LSTMs to model the temporal patterns in the data, facilitating quantum state diffusion simulations.

\begin{figure*}
    \centering
    \includegraphics[scale=0.45]{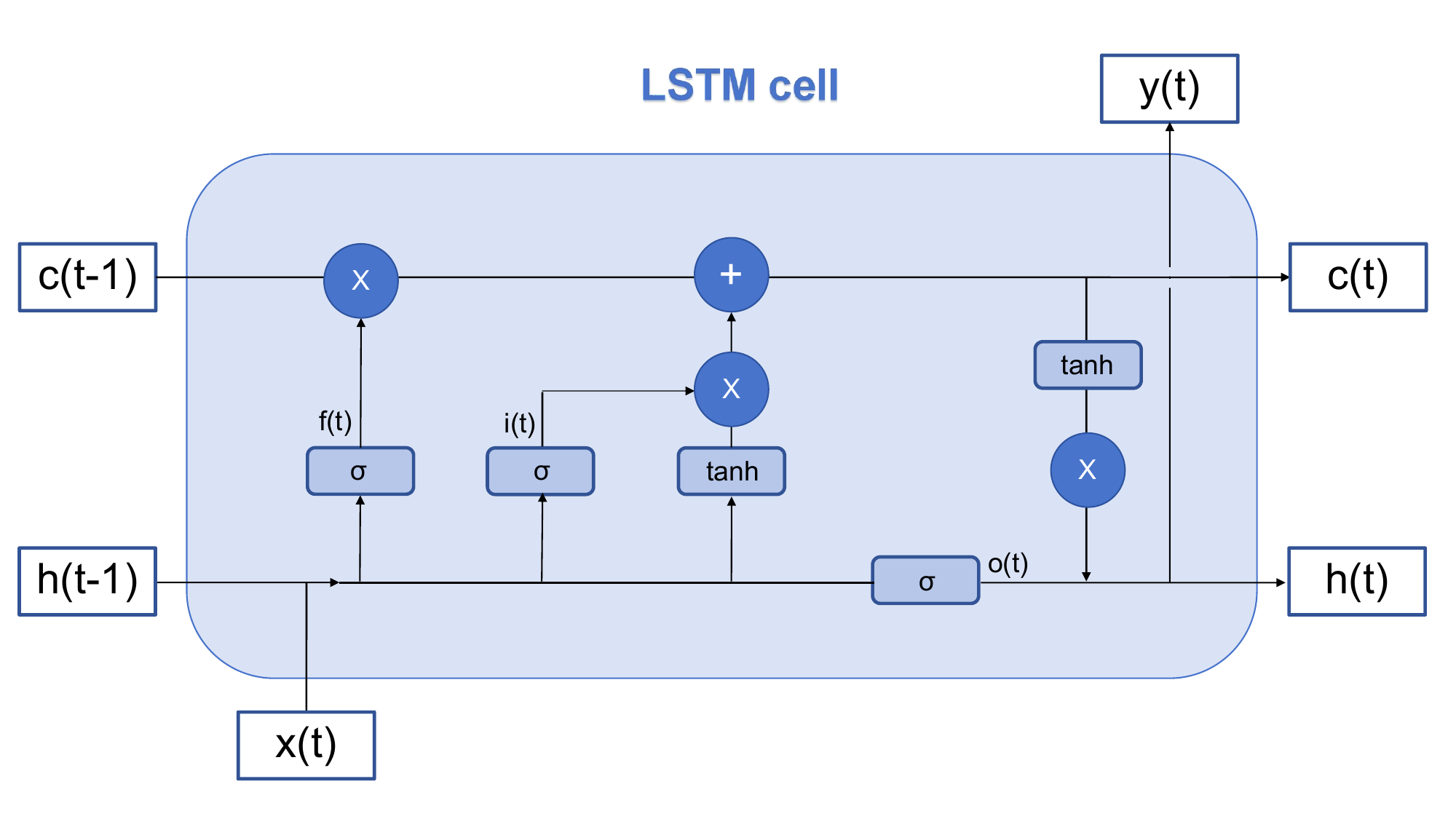}
    \caption{The structure of the LSTM cell. $\boldsymbol{\sigma}$ denotes the sigmoid activation function. $\mathbf{tanh}$ denotes hyperbolic tangent activation function. $\oplus$ denotes the broadcasting addition and $\otimes$ denotes the element-wise multiplication.}
    \label{fig:lstmstr}
\end{figure*}

The architecture of the LSTMs' unit, termed as LSTM cell, is shown in Figure~\ref{fig:lstmstr}, which can be simply understood as a function:
\begin{equation}
    (\boldsymbol{y}_{(t)},\boldsymbol{c}_{(t)})=(\boldsymbol{h}_{(t)},\boldsymbol{c}_{(t)})=f(\boldsymbol{x}_{(t)},\boldsymbol{c}_{(t-1)},\boldsymbol{h}_{(t-1)})
\end{equation}
For each cell, the vector $\boldsymbol{x}_{(t)}$ is the input vector and $\boldsymbol{y}_{(t)}$ is the output vector. The vectors $\boldsymbol{c}_{(t-1)}$ and $\boldsymbol{h}_{(t-1)}$ are the vectors carrying historical information from the previous LSTM cell, respectively termed as the long-term state vector and short-term state vector.

The LSTM cell receives vector $\boldsymbol{x}_{(t)}$ at step t and the state vectors $\boldsymbol{c}_{(t-1)}$ and $\boldsymbol{h}_{(t-1)}$ from the last step $t-1$, and then generates output vector $\boldsymbol{y}(t)$ by processing them through three gates\cite{lstmgates}: input gate, forget gate and output gate. The mathematical descriptions are as follows:

\begin{equation}
    \begin{aligned}
&\boldsymbol{i}_{(t)}=\sigma(\boldsymbol{W}_{xi}^{T}\boldsymbol{x}_{(t)}+\boldsymbol{W}_{hi}^{T}\boldsymbol{h}_{(t-1)}+\boldsymbol{b}_{i}) \\
&\boldsymbol{f}_{(t)}=\sigma(\boldsymbol{W}_{xf}^{T}\boldsymbol{x}_{(t)}+\boldsymbol{W}_{hf}^{T}\boldsymbol{h}_{(t-1)}+\boldsymbol{b}_{f}) \\
&\boldsymbol{o}_{(t)}=\sigma(\boldsymbol{W}_{x\sigma}^{T}\boldsymbol{x}_{(t)}+\boldsymbol{W}_{ho}^{T}\boldsymbol{h}_{(t-1)}+\boldsymbol{b}_{o}) \\
&\boldsymbol{g}_{(t)}=\operatorname{tanh}(\boldsymbol{W}_{xg}^{T}\boldsymbol{x}_{(t)}+\boldsymbol{W}_{hg}^{T}\boldsymbol{h}_{(t-1)}+\boldsymbol{b}_{g}) \\
&\boldsymbol{c}_{(t)}=\boldsymbol{f}_{(t)}\cdot \boldsymbol{c}_{(t-1)}+\boldsymbol{i}_{(t)}\cdot \boldsymbol{g}_{(t)} \\
&\boldsymbol{y}_{(t)}=\boldsymbol{h}_{(t)}=\boldsymbol{o}_{(t)}\cdot\mathrm{tanh}(\boldsymbol{c}_{(t)})
\end{aligned}
\end{equation}
where $\sigma$ denotes the sigmoid activation function and $\mathbf{tanh}$ denotes hyperbolic tangent activation function. The input gate decides the relevant part of the input to be processed. The forget gate regulates the amount of information to be discarded. The output gate then produces the output vector $\boldsymbol{o}_{(t)}$ and forwards it to the next cell as the short-term vector $\boldsymbol{h}_{(t)}$.

Effective LSTMs are typically constructed by connecting several LSTM cells in a multi-layer structure. 

\subsubsection{Attentional Feature Fusion}

Feature fusion is another important technique extensively applied in various deep learning architectures to combine features from different layers, models, and sources. Traditional methods of feature fusion include concatenation, element-wise addition, and multiplication. As machine learning has advanced, researchers have developed more sophisticated strategies for feature fusion. Kernel-based fusion methods, such as Kernel Principal Component Analysis (KPCA)\cite{kpca} and Support Vector Machines (SVMs)\cite{svm} with multiple kernels, allow for the non-linear combination of features. Ensemble methods like bagging and boosting combine outputs from multiple base learners to achieve more accurate predictions. With the advancement of deep learning techniques, particularly CNNs and RNNs, feature fusion methods that leverage hierarchical features extracted from raw data have been developed. Examples include skip connections in Residual Networks (ResNets)\cite{res} and multi-scale fusion strategies in Feature Pyramid Networks (FPNs)\cite{fpn} and U-Nets\cite{unet}. Recently, attention mechanisms have been integrated into feature fusion methods. Models based on Transformer architecture, for instance, use self-attention and multi-head attention mechanisms. Additionally, Squeeze-and-Excitation Networks (SENet)\cite{senet} adaptively recalibrate channel-wise feature responses. In this work, we employ iterative attentional feature fusion (iAFF)\cite{aff}, a unified and versatile approach, to integrate the various features extracted from the quantum dynamical evolution.

Attentional feature fusion is based on the multi-scale channel attention module (MS-CAM).
Given the intermediate feature $\boldsymbol{X}$, MS-CAM provide the global channel context $\boldsymbol{G}(\boldsymbol{X})$ and the local channel context $\boldsymbol{L}(\boldsymbol{X})$.
\begin{figure*}
    \centering
    \includegraphics[scale=0.45]{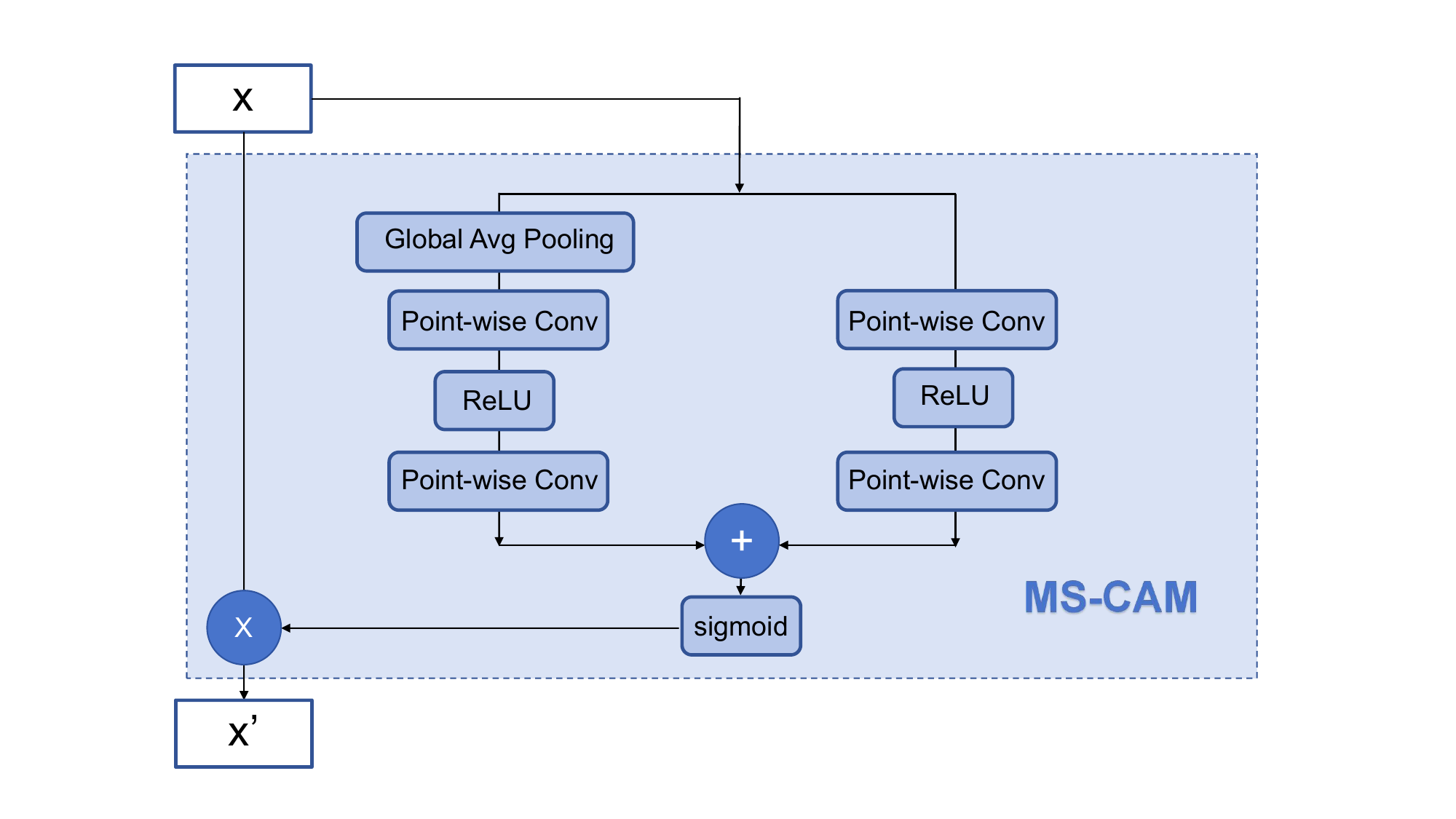}
    \caption{The structure of the multi-scale channel attention module.}
    \label{fig:ms-cam}
\end{figure*}
By varying the spatial pooling size, channel attention can be applied at multiple scales. Point-wise convolution (PWConv) aggregates local channel context, while global average pooling (GlobalAvgPooling) encodes global features by averaging across each channel.
\begin{equation}
    \boldsymbol{G}(\boldsymbol{X})=\mathcal{B}\left(\text{PWConv}_2\left(\mathcal{\delta}\left(\mathcal{B}\left(\text{PWConv}_1(\text{GlobalAvgPooling}(\boldsymbol{X}))\right)\right)\right)\right)
\end{equation}
\begin{equation}
    \boldsymbol{L}(\boldsymbol{X})=\mathcal{B}\left(\text{PWConv}_4\left(\mathcal{\delta}\left(\mathcal{B}\left(\text{PWConv}_3(\boldsymbol{X})\right)\right)\right)\right)
\end{equation}
where $\mathcal{B}$ denotes the Batch Normalization (BN). $\mathcal{\delta}$ is the activation function Rectified Linear Unit (ReLU).
Next, MS-CAM obtains the refined feature $\boldsymbol{X'}$ as follow:
\begin{equation}
    \boldsymbol{X'}=\boldsymbol{X}\otimes\mathcal{M}(\boldsymbol{X})=\boldsymbol{X}\otimes\sigma\left(\boldsymbol{L}(\boldsymbol{L})\oplus\boldsymbol{g}(\boldsymbol{X})\right)
\end{equation}
where $\oplus$ denotes the broadcasting addition and $\otimes$ denotes the element-wise multiplication. 

\begin{figure*}
    \centering
    \includegraphics[scale=0.45]{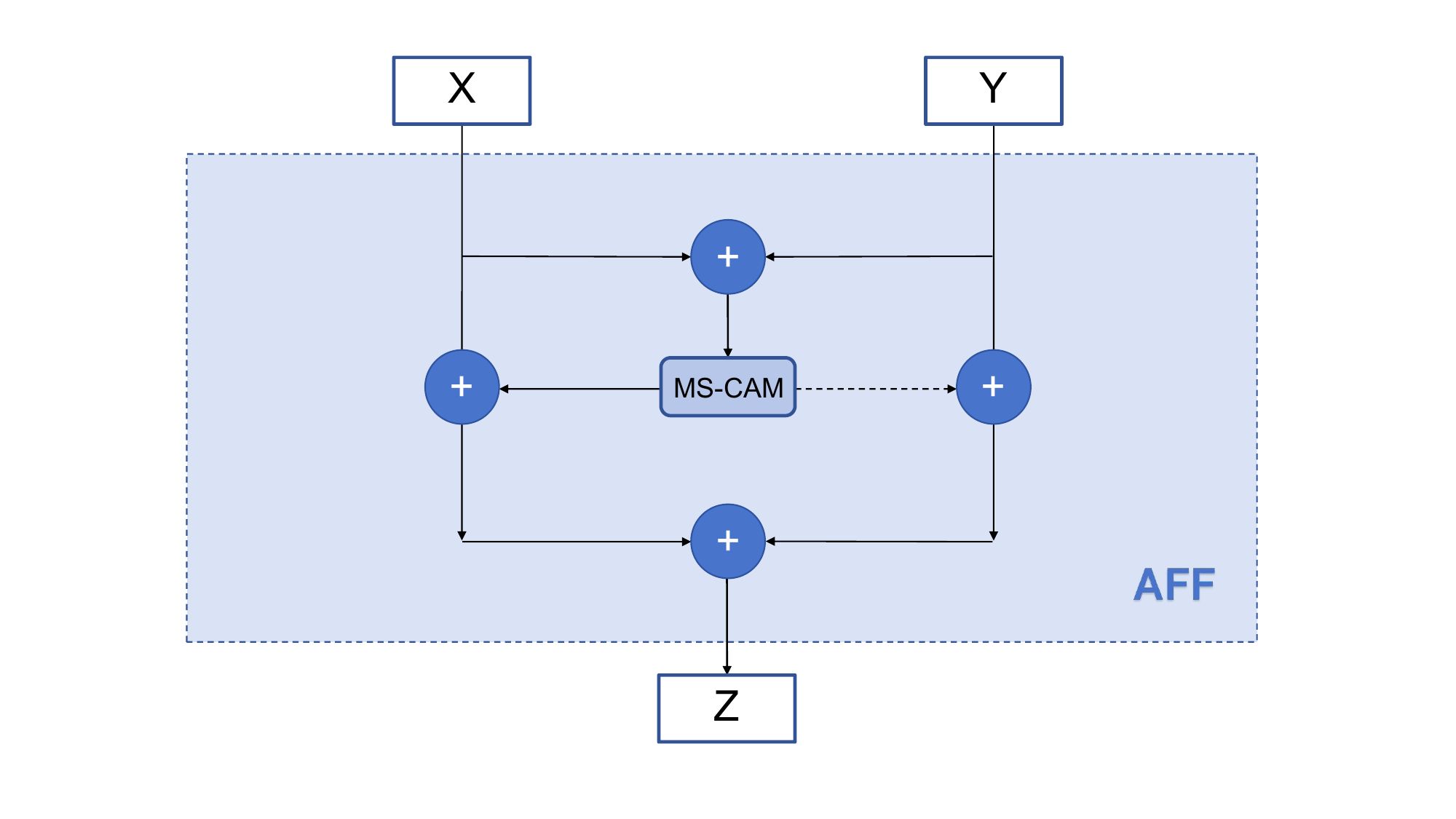}
    \caption{The structure of attentional feature fusion (AFF).}
    \label{fig:aff}
\end{figure*}

Based on the multi-scale channel attention module $\mathcal{M}$, attentional feature fusion structure can be introduced:
\begin{equation}
    \boldsymbol{Z}=\mathcal{M}(\boldsymbol{X}\uplus \boldsymbol{Y})\otimes \boldsymbol{X}+(1-\mathcal{M}(\boldsymbol{X}\uplus \boldsymbol{Y}))\otimes \boldsymbol{Y}
\end{equation}

\begin{figure*}
    \centering
    \includegraphics[scale=0.45]{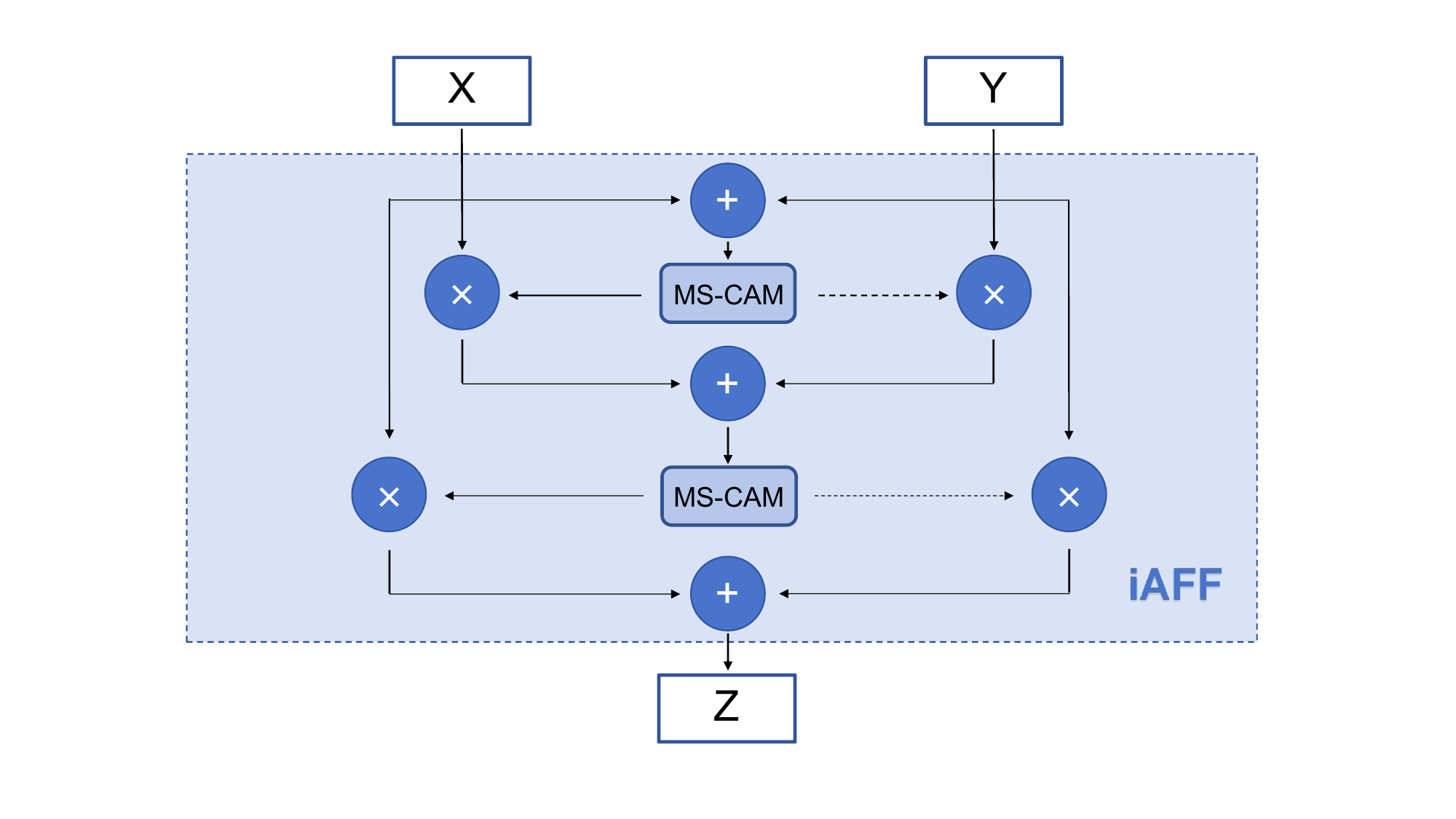}
    \caption{The structure of iterative attentional feature fusion. The initial feature integration is implemented as the Attentional Feature Fusion (AFF) module.}
    \label{fig:iaff}
\end{figure*}

For simplicity, the initial feature integration can be implemented as an element-wise summation. However, that may not be optimal. A more effective strategy is to introduce an additional attention module to fuse the input features. By replacing the initial feature integration with the AFF module, we obtain an iterative process denoted as iterative attentional feature fusion.

\begin{equation}
    \boldsymbol{X}\uplus\boldsymbol{Y}=\mathcal{M}(\boldsymbol{X}+\boldsymbol{Y})\otimes\boldsymbol{X}+(1-\mathcal{M}(\boldsymbol{X}+\boldsymbol{Y}))\otimes\boldsymbol{Y}
\end{equation}

\subsection{\label{sec:scheme}3.The Deep Learning Scheme for Quantum State Diffusion Simulations}

\subsubsection{3.1 cNMSSE Simulation and Convergence Assessment}

In the forward and backward cNMSSEs simulation, the raw data consists of the reduced density matrices, which are obtained by ensemble averaging the outer products of the forward and backward stochastic state vectors.
\begin{equation}
    \boldsymbol{\bar\rho}=\sum^N_{n=1}\frac{\boldsymbol{\rho_n}}{N}=\sum^N_{n=1}\frac{|\boldsymbol{\psi}^f_n\rangle\langle\boldsymbol{\psi}^b_n|}{N}
\end{equation}

To analyze the convergence of stochastic trajectories, we first simulate \( N \) stochastic trajectories, denoted as $\{\bm{\rho}_n(t_1),\bm{\rho}_n(t_2),\bm{\rho}_n(t_3),\cdots,\bm{\rho}_n(t_T)\}_{n=1}^{N}$ using the cNMSSEs. We then calculate the mean trajectory values $\{\bm{\bar\rho}(t_1),\bm{\bar\rho}(t_2),\bm{\bar\rho}(t_3),\cdots,\bm{\bar\rho}(t_T)\}$ and the standard error (SE) $\{\boldsymbol{S}(t_1),\boldsymbol{S}(t_2),\boldsymbol{S}(t_3),\cdots,\boldsymbol{S}(t_T)\}$ across all trajectories. Based on the standard error values, the propagation be segmented into two parts: the converged part $\{\bm{\bar\rho}(t_1),\bm{\bar\rho}(t_2),\cdots,\bm{\bar\rho}(t_c)\}$,where $\{\boldsymbol{S}(t_i)\}^{c}_{i=1}\leq\epsilon_1$, and the oscillating part $\{\bm{\bar\rho}(t_{c+1}),\bm{\bar\rho}(t_{c+2}),\cdots,\bm{\bar\rho}(t_T)\}$. The standard error quantifies the convergence of random trajectories. We define $\epsilon_1$ as the set threshold for acceptable accuracy. A smaller $\epsilon_1$ reduces the oscillations in the training data, resulting in more precise final predictions. However, achieving this increased accuracy requires a greater number of random trajectories, leading to higher computational costs.

Both the converged and oscillating sections are utilized for the initial pre-training of the neural network. However, only the converged section is used for fine-tuning. In our numerical experiments, pre-training with the entire dataset, including the oscillating section, significantly enhanced prediction accuracy. While these results may still exhibit oscillations, particularly over long-term evolution, they retain valuable dynamical information related to the steady state, even if partially obscured by noise.

\subsubsection{3.2 Dataset Preparation}

Before sending the reduced density matrices $\{\boldsymbol{\bar\rho}_t\}^T_{t=1}=\{\boldsymbol{\bar\rho}(t_1),\boldsymbol{\bar\rho}(t_2),\cdots,\boldsymbol{\bar\rho}(t_T)\}$ to the neural networks, we convert them into vectors labeled as $\{\boldsymbol{\omega}_t\}^T_{t=1}=\{\boldsymbol{\omega}(t_1),\boldsymbol{\omega}(t_2),\cdots,\boldsymbol{\omega}(t_T)\}$. This transformation removes unnecessary terms and retains only the essential information. For the diagonal elements, only the population difference is preserved, as demonstrated in Eq.~(\ref{vec}):

\begin{equation}
    \\\Bigg(\begin{matrix}\rho_{11} \ldots \rho_{1n}\\ 
\vdots \\\rho_{n1}\ldots \rho_{nn}\end{matrix}\Bigg)\rightarrow\begin{pmatrix}\Delta_{1}\\\vdots\\\Delta_{n-1}\\\vdots \\ \mathrm{Re}\{\rho_{1~2}\}\\ \mathrm{Im}\{\rho_{1~2}\}\\\vdots\\\mathrm{Re}\{\rho_{n-1~n}\}\\\mathrm{Im}\{\rho_{n-1~n}\}\end{pmatrix}\label{vec}
\end{equation}
where $\Delta_1=\rho_{11}-\rho_{22},\Delta_2=\rho_{22}-\rho_{33},...,\Delta_{n-1}=\rho_{n-1~n-1}-\rho_{n~n}$. $\mathrm{Re\{\rho_{ij}\}_{i<j}}$ and $\mathrm{Im\{\rho_{ij}\}_{i<j}}$ is the real and imaginary part of the off-diagonal element. 

For the off-diagonal elements, both their real and imaginary parts will be preserved. Since the reduced density matrix is Hermitian, the real part of the off-diagonal element and the real part of its symmetric term are the same, while their imaginary part are inverse. Therefore, there is no need to reintroduce redundant information to the vector.
The time series neural networks often take multiple input vectors $\{\boldsymbol{\omega}(t_a),\boldsymbol{\omega}(t_{a+1}),...,\boldsymbol{\omega}(t_{a+L-1})\}$, and predict a single output vector $\boldsymbol{\omega}(t_{a+L})$.
The number $L$ of input vectors significantly impacts the predictive performance. $L$ vectors continuing in time will be packed into one group labeled as $\boldsymbol{x}(t_a)=\{\boldsymbol{\omega}(t_a),\boldsymbol{\omega}(t_{a+1}),...,\boldsymbol{\omega}(t_{a+L-1})\}$, which serves as an input to the deep neural networks. The model's ability to capture dynamic correlation features varies with changes in the size of the input vector.

The entire evolution period is divided into three parts: training set, validation set and prediction zone. Data $\{\boldsymbol{\omega}_t\}^c_{t=1}=\{\boldsymbol{\omega}(t_1),\boldsymbol{\omega}(t_2),\cdots,\boldsymbol{\omega}(t_c)\}$ from cNMSSEs with standard error (SE) below $\epsilon_1$ can be selected as the converged data. And the $\{\boldsymbol{x}_t\}^{c-L+1}_{t=1}=\{\boldsymbol{x}(t_1),\boldsymbol{x}(t_2),\cdots,\boldsymbol{x}(t_{c-L+1})\}$ made from $\{\boldsymbol{\omega}_t\}_{t=1}^{c}$ can be selected for use in the training and validation sets. 

We first divide $\{\boldsymbol{x}_t\}_{t=1}^{c-L+1}$ into two groups with a ratio of 3:1, resulting in a set \(\mathbf{T}\) and a validation set \(\mathbf{V}\). Then \(\mathbf{T}\) is randomly shuffled. Subsequently, we partitioned \(\mathbf{T}\) into two groups with a ratio of 7:3, resulting in a set \(\mathbf{T}_1\) and a set \(\mathbf{T}_2\). The model is trained using the set \(\mathbf{T}_1\). To prevent overfitting, we employed an early stopping strategy based on the model's performance on the test set \(\mathbf{T}_2\). The training set \(\mathbf{T}\) offers accurate correlation information and is used to train the neural networks for predicting the long-term behavior of the quantum system. The validation set, denoted as \(\mathbf{V}\), is used to select the optimal hyperparameters. The trained model, with various structures, is tested on \(\mathbf{V}\). The model that yields the lowest error will be used to predict the long-term evolution.

\subsubsection{3.3 Model construction and Grid Searching}

\begin{figure*}
    \centering
    \includegraphics[scale=0.45]{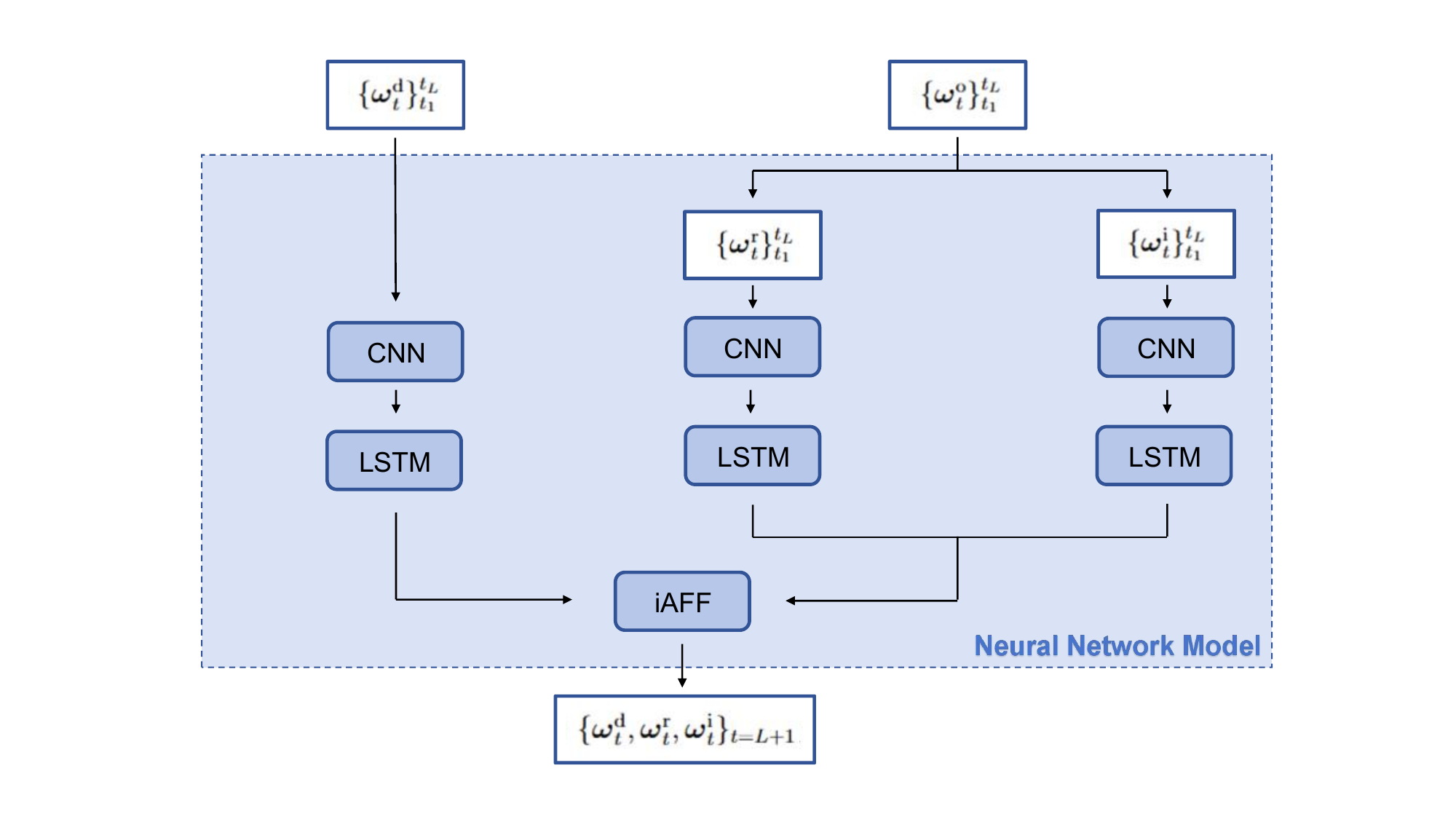}
    \caption{The overall structure of the neural network model.}
    \label{fig:neural_network_model}
\end{figure*}

The proposed neural network structure comprises CNNs, LSTMs, and iAFF components. As illustrated in Figure~\ref{fig:neural_network_model}, the input data $\{\boldsymbol{\omega}_t\}_{t_1}^{t_L}=\{\boldsymbol{\omega}({t_1}),\boldsymbol{\omega}({t_2}),\cdots,\boldsymbol{\omega}({t_L})\}$ is separated into two components: the diagonal elements $\{\boldsymbol{\omega}_t^{\text{d}}\}_{t_1}^{t_L}=\{\boldsymbol{\omega}({t_1})^{\text{d}},\boldsymbol{\omega}({t_2})^{\text{d}},\cdots,\boldsymbol{\omega}({t_L})^{\text{d}}\}$, representing the dynamic evolution, and the off-diagonal elements $\{\boldsymbol{\omega}_{t}^{\text{o}}\}_{t_1}^{t_L}=\{\boldsymbol{\omega}({t_1})^{\text{o}},\boldsymbol{\omega}({t_2})^{\text{o}},\cdots,\boldsymbol{\omega}({t_L})^{\text{o}}\}$, which encode coherent information. Given that the off-diagonal elements are complex, $\{\boldsymbol{\omega}_{t}^{\text{o}}\}_{t_1}^{t_L}$ are separated into their real $\{\boldsymbol{\omega}_{t}^{\text{r}}\}_{t_1}^{t_L}$ and imaginary $\{\boldsymbol{\omega}_{t}^{\text{i}}\}_{t_1}^{t_L}$ parts.

The CNNs are used to extract local features separately from the diagonal elements, the real parts of the off-diagonal elements, and the imaginary parts of the off-diagonal elements. These extracted features are then processed by three sets of LSTM layers to identify intrinsic dynamic correlations. The outputs from the real and imaginary parts are combined. Finally, the features from both the diagonal and off-diagonal components are integrated using iAFF. The neural networks outputs the vector $\{\boldsymbol{\omega}_{t}^{\text{d}},\boldsymbol{\omega}_{t}^{\text{r}},\boldsymbol{\omega}_{t}^{\text{i}}\}_{t=L+1}$ for the next time step.

In the neural network model, several hyperparameters need to be defined: the size $p$ and stride $r$ of the convolutional kernel in the CNNs, the number of neurons in the LSTM layer, and the input length $L$ for the time series. To determine the optimal configuration, we perform testing across these hyperparameters.

In the numerical experiments, We construct models with various topologies, differing in kernel size, numbers of neurons, and input lengths. Detailed settings are given in appendix. These models are trained on the training set $\mathbf{T}$ and evaluated on the validation set $\mathbf{V}$. The model that achieves the lowest validation error is chosen for future forecasting of system evolution.

\subsubsection{3.4 Pre-train and Fine tuning}

The entire period of evolution data from cNMSSEs can be utilized as a pre-training set. The model with optimal hyperparameters trained by this pre-training set can provide approximately correct predictions. Next, the dataset $\mathbf{T}$, filtered to include samples with standard error below a threshold $\epsilon_1$, is used for fine-tuning. We aim to refine the model without fully retraining it. Thus, the number of training epochs and the learning rate are reduced to prevent overfitting.

During training, the model's weights and biases are adjusted to reduce the difference between the original data and the generated output. This process ensures that the temporal patterns within the sequence data are accurately captured. Once trained, the neural network can be used to model future time series as non-linear dynamic systems.

\subsubsection{3.5 Predicted results}

A major challenge in many neural network forecasting tasks is assessing the accuracy of predictions, which often lack reference points for validation. The neural network-based NMSSE method addresses this issue by ensuring inherent convergence. This method incorporates a feedback mechanism that systematically validates the accuracy of predictions, enhancing reliability for extended forecasting.

We begin by generating 10 sets of nmsse simulations $\{\bm{\rho}_n(t)\}_{n=1}^{N_1},\{\bm{\rho}_n(t)\}_{n=1}^{N_2},\cdots,\{\bm{\rho}_n(t)\}_{n=1}^{N_{10}}$ with increasing trajectory number from $N_1$ to $N_{10}$. Each set is used to train a potential dynamic map. As the number of trajectories increases, the predictions $\{\boldsymbol{\hat\rho}(t)\}^{N_1},\{\boldsymbol{\hat\rho}(t)\}^{N_2},\cdots,\{\boldsymbol{\hat\rho}(t)\}^{N_{10}}$ become more stable and converged. To assess this, we calculate the standard deviation (SD) across these 10 predictions. If the SD is below a threshold value, $\epsilon_2$, and the result from the tenth prediction is accepted as the final outcome. However, if the SD exceeds $\epsilon_2$, additional stochastic trajectories are required from NMSSE simulation. Reducing the value of $\epsilon_2$ improves accuracy but also increases computational cost.

Adopting neural network techniques to enhance the non-Markovian stochastic Schrödinger equations simulation can utilize the inherent compatibility between NMSSE and neural network approaches. Neural networks are highly effective at pattern recognition and extracting essential information, while NMSSE provides evolving temporal data often obscured by noise. One of the primary challenges in applying neural networks to time series prediction is the difficulty of validating accuracy when future data is unavailable for reference. In contrast, the NMSSE method converges with an increasing number of trajectories, enabling accuracy assessment through the standard deviation of predictions derived from expanding data. While NMSSE can experience oscillations during long-term evolution, neural networks, trained as nonlinear dynamic maps, can effectively manage stable long-term propagation by leveraging the converged short-term data from NMSSE. In principle, combining neural networks and NMSSE could address the convergence challenges, reduce the required number of stochastic trajectories, and offer a reliable standard for predictive accuracy. 

\section{Numerical Results and Discussion}

To validate the robustness of the proposed NN-NMSSE scheme, this section presents two benchmark simulations, quantum population dynamics in the Spin-Boson Model (SBM) and energy transfer in the Fenna-Matthews-Olson (FMO) complex.

In the following calculations, the cNMSSEs Eq.~(\ref{eq:cnmsse}) are applied to obtain the raw data for neural network training. The proposed networks, which integrates CNNs, LSTMs, and iAFF components, is applied to process the long-term evolution. Through an iterative process, the standard deviation of predictions, used as convergence criteria, gradually decrease until they reach the predefined threshold $\epsilon_2$, at which point the results are considered converged.

\subsection{1.The Spin-Boson Model}

We begin by applying the cNMSSE to simulate the quantum dynamics of the Spin-Boson Model (SBM) \cite{sbm}.

The Hamiltonian for the two-level system is defined as $\hat{H}_{S}=\epsilon\hat{\sigma}_{z}+{V\hat{\sigma}}_{x}$, where $\hat{\sigma}_{x}=|1\rangle\langle2|+|2\rangle\langle1|$ and $\hat{\sigma}_{z}=|1\rangle\langle1|-|2\rangle\langle2|$. For the simulations, we set $\epsilon = 1.0$ and the electronic coupling \(V\) = 1.0. In this model, fluctuations in site energy arise from linear electron-phonon interactions, with phonon properties characterized by the spectral density function. We employ the Debye-Drude spectral density, given by $J(\omega)=\frac{\eta\omega\gamma}{\omega^{2}+\gamma^{2}}$, where the system-bath coupling strength $\eta$ is set to 0.5. Identified by the ratio $\frac{\gamma}{V}$, the SBM dynamics can be classified as the adiabatic  and the non-adiabatic  regimes. To explore both regimes, we vary the phonon characteristic frequency $\gamma$ as $0.25, 1.0$, and $5.0$, spanning adiabatic $(\frac{\gamma}{V} < 1)$ to non-adiabatic $(\frac{\gamma}{V} > 1)$ behavior. Because the charge carrier's coherent motion is sensitive to temperature and exhibits oscillatory behavior at low temperatures, we test the proposed neural network-based NMSSE scheme at both high temperature ($\beta=0.5$) and low temperature ($\beta=5.0$). Accordingly, we validate the NN-NMSSE method across various parameter sets: (a) $\beta = 5.0$, $\gamma = 5.0$; (b) $\beta = 5.0$, $\gamma = 1.0$; (c) $\beta = 5.0$, $\gamma = 0.25$; (d) $\beta = 0.5$, $\gamma = 5.0$; (e) $\beta = 0.5$, $\gamma = 1.0$; (f) $\beta = 0.5$, $\gamma = 0.25$.

\begin{figure}
    \centering   \includegraphics[scale=0.5]{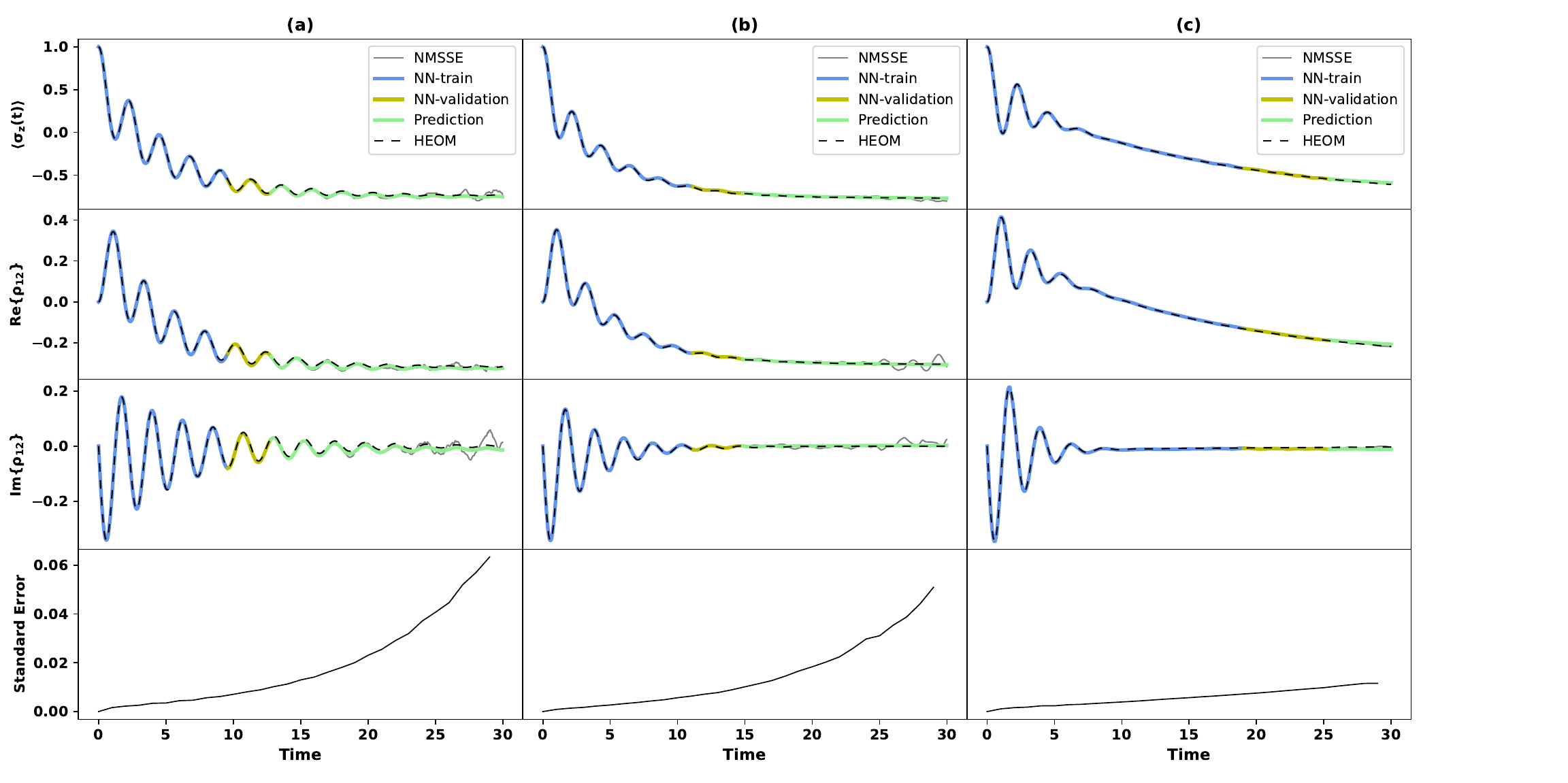}
    \caption{Compare the NN-NMSSE predictions, shown as colored lines, with the exact results (black dashed lines) derived from the HEOM method at low temperature ($\beta = 5.0$). The comparisons are presented for different parameter sets: (a) $\beta = 5.0$, $\gamma = 5.0$; (b) $\beta = 5.0$, $\gamma = 1.0$; (c) $\beta = 5.0$, $\gamma = 0.25$. The HEOM results, represented by black dashed lines, serve as the reference for evaluating the accuracy of the NN-NMSSE approach under each parameter condition. The results of NMSSE simulations using the same number of stochastic trajectories as NN-NMSSE, indicated by gray lines, are presented. In these results, oscillations are observed in the long-time regime. The figures in the fourth row illustrate the standard error of $\langle\sigma_z(t)\rangle$ across the NMSSE stochastic trajectories. The temporal pattern of this standard error determines the extent to which the NN-NMSSE approach can enhance the original method.}
    \label{fig:low1}
\end{figure}

Figure~\ref{fig:low1} presents NMSSE simulations and NN-NMSSE predictions at low temperature ($\beta=5.0$). As shown, the NN-NMSSE method effectively learns the intrinsic propagation pattern from the oscillatory raw data and successfully predicts the long-term evolution with minimal oscillations. The robustness of NN-NMSSE is also demonstrated across both adiabatic and non-adiabatic regimes.

\begin{figure}
    \centering  \includegraphics[scale=0.5]{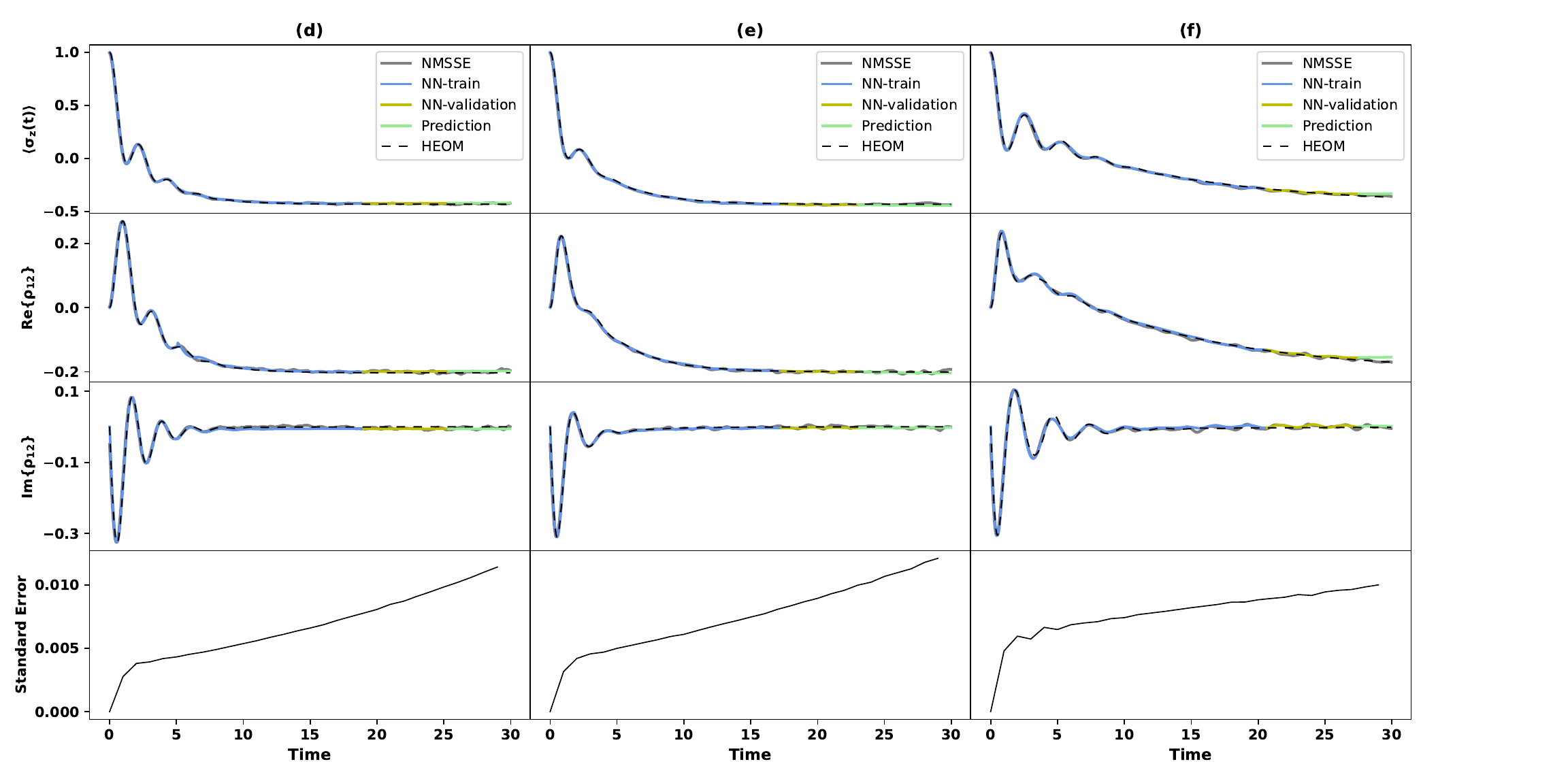}
    \caption{Compare the NN-NMSSE predictions, shown as colored lines, with the exact results (black dashed lines) derived from HEOM method at high temperature ($\beta = 0.5$). The comparisons are presented for different parameter sets: (d) $\beta = 0.5$, $\gamma = 5.0$; (e) $\beta = 0.5$, $\gamma = 1.0$; (f) $\beta = 0.5$, $\gamma = 0.25$. The HEOM results, represented by black dashed lines, serve as the reference for evaluating the accuracy of the NN-NMSSE approach under each parameter condition.}
    \label{fig:high1}
\end{figure}

In this experiment, we set the standard error threshold $\epsilon_1$ of the trajectories to 0.01 and the standard deviation threshold $\epsilon_2$ for predictions to 0.05. As noted in the Methodology, using a smaller threshold reduces oscillations in the raw data, leading to more accurate predictions. However, this improvement comes at the cost of increased computational demand. These settings were consistently applied in the subsequent high-temperature experiment (shown in Figure~\ref{fig:high1}).

At low temperature and non-adiabatic regime, short-term evolution exhibits significantly reduced oscillations compared to long-term evolution (as shown in the fourth row of Figure~\ref{fig:low1}). Consequently, neural networks trained on short-term data can produce long-term predictions with fewer oscillations than those generated by NMSSE simulations. This highlights the NN-NMSSE method’s capacity to manage oscillation issue in low-temperature non-Markovian simulations.

At high temperature, however, minor oscillations remain in the long-term predictions of the NN-NMSSE method. This occurs because the standard error of the stochastic trajectories varies slightly over time at high temperature, leading to comparable levels of oscillations in both short- and long-term evolution (as shown in the fourth row of Figure~\ref{fig:high1}). As a result, the neural networks can only capture oscillatory patterns from short-term data, limiting its ability to predict smooth long-term dynamics.

It’s worth noting that NMSSE performs well at high temperature and adiabatic regime, making the NN-NMSSE approach unnecessary in these conditions. This will be further illustrated in the following cost comparison.

\begin{figure}
    \includegraphics[scale=0.5]{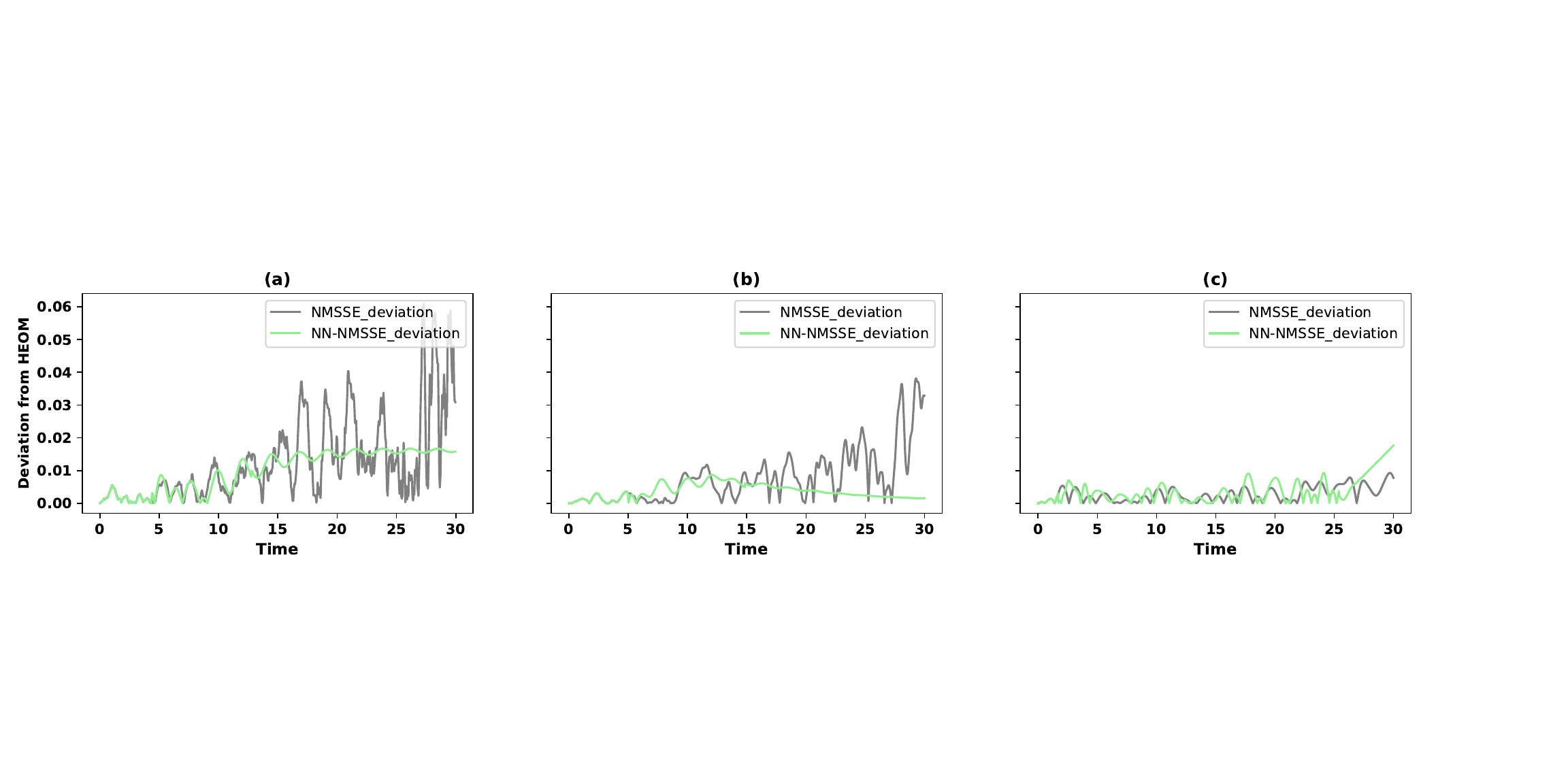}
    \caption{ 
    We compares the performance of NMSSE and NN-NMSSE method using the same number of stochastic trajectories across three low-temperature experiments: (a) $\beta = 5.0$, $\gamma = 5.0$ with 180,000 trajectories; (b) $\beta = 5.0$, $\gamma = 1.0$ with 540,000 trajectories; (c) $\beta = 5.0$, $\gamma = 0.25$ with 150,000 trajectories. The figure illustrates the population deviations between two methods and the exact results from the HEOM benchmark. The deviations for the NN-NMSSE approach are represented by green lines, while those for the NMSSE simulation method are shown in gray. The results highlight the advantages of the NN-NMSSE method, particularly in scenarios where the NMSSE method encounters significant oscillations during long-term simulations.}
    \label{fig:low_de}
\end{figure}

Using the NN-NMSSE method, the deviation between its predictions and the HEOM benchmark remained below approximately 0.02 (as shown in Figure~\ref{fig:low_de}). In comparison, achieving the same level of accuracy with NMSSE required a much more computational cost (as shown in Figure~\ref{fig:cost-time}). This result demonstrates that NN-NMSSE significantly reduces the number of trajectories and computational cost needed for accurate predictions at low temperature.

\begin{figure}
    \includegraphics[scale=0.45]{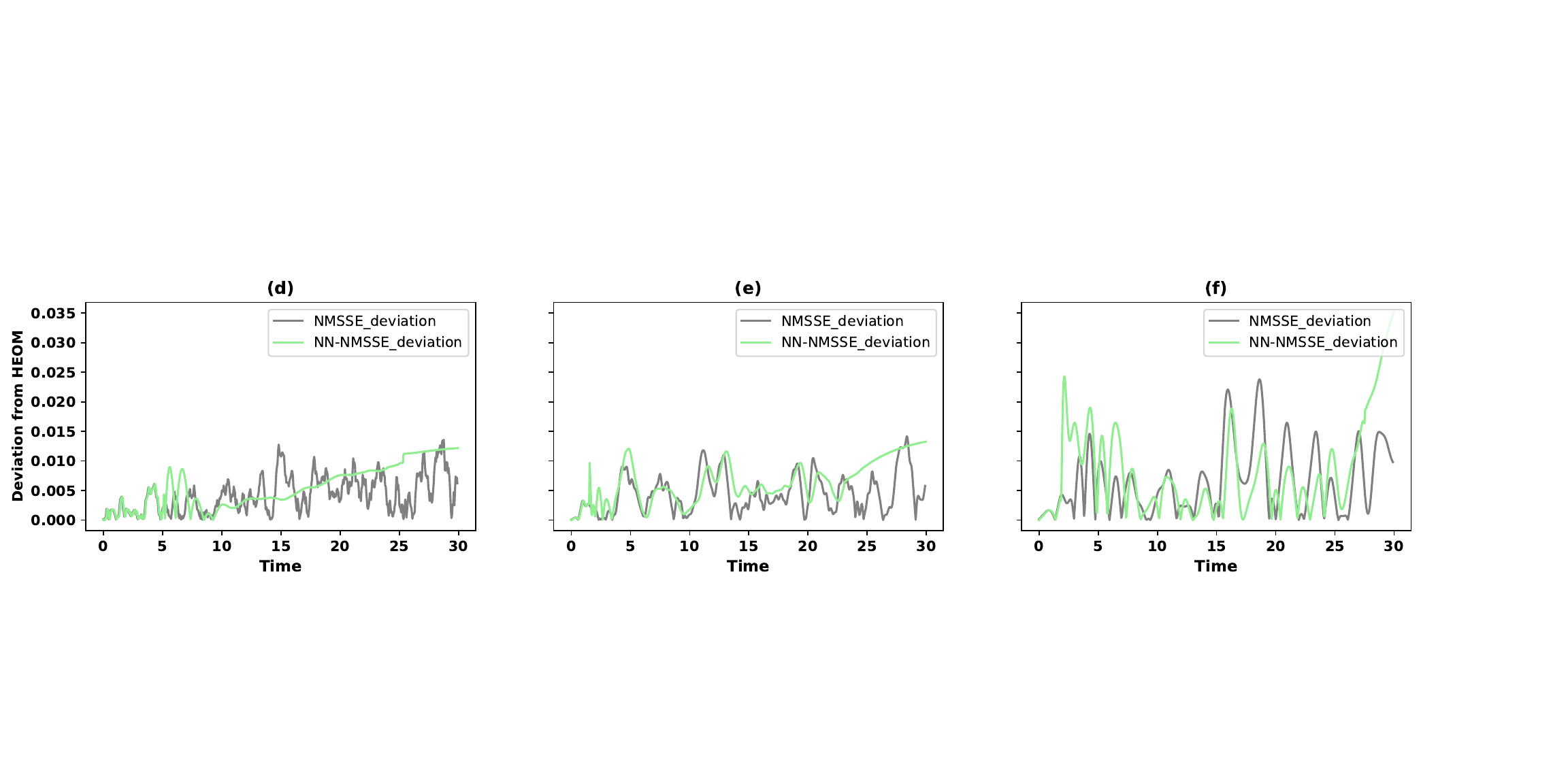}
    \caption{The figure illustrates the population deviations between the two methods and the exact results from the HEOM benchmark. The deviations for the NN-NMSSE approach are represented by green lines, while those for the NMSSE simulation method are shown in gray. Three high-temperature experiments are presented: (d) $\beta = 0.5$, $\gamma = 5.0$ with 39,000 trajectories; (e) $\beta = 0.5$, $\gamma = 1.0$ with 28,000 trajectories; (f) $\beta = 0.5$, $\gamma = 0.25$ with 10,000 trajectories. Compared to simulations at low temperatures, the NMSSE method can achieve convergence with significantly fewer trajectories at highe temperatures.}
    \label{fig:high_de}
\end{figure}

Figure~\ref{fig:cost} and Figure~\ref{fig:cost-time} further summarize the the trajectory counts and computational time required across all experiments. Notably, at low temperatures, NMSSE simulations require far more stochastic trajectories to achieve convergence than in high-temperature scenarios. The NN-NMSSE method effectively addresses these convergence challenges, reducing the required number of trajectories. At high temperatures, the NN-NMSSE method does not enhance NMSSE method. However, NMSSE simulations at high temperatures do not encounter convergence issues. In these conditions, NMSSE achieves accurate results with a relatively low number of trajectories.

\begin{figure}
    \includegraphics[scale=0.45]{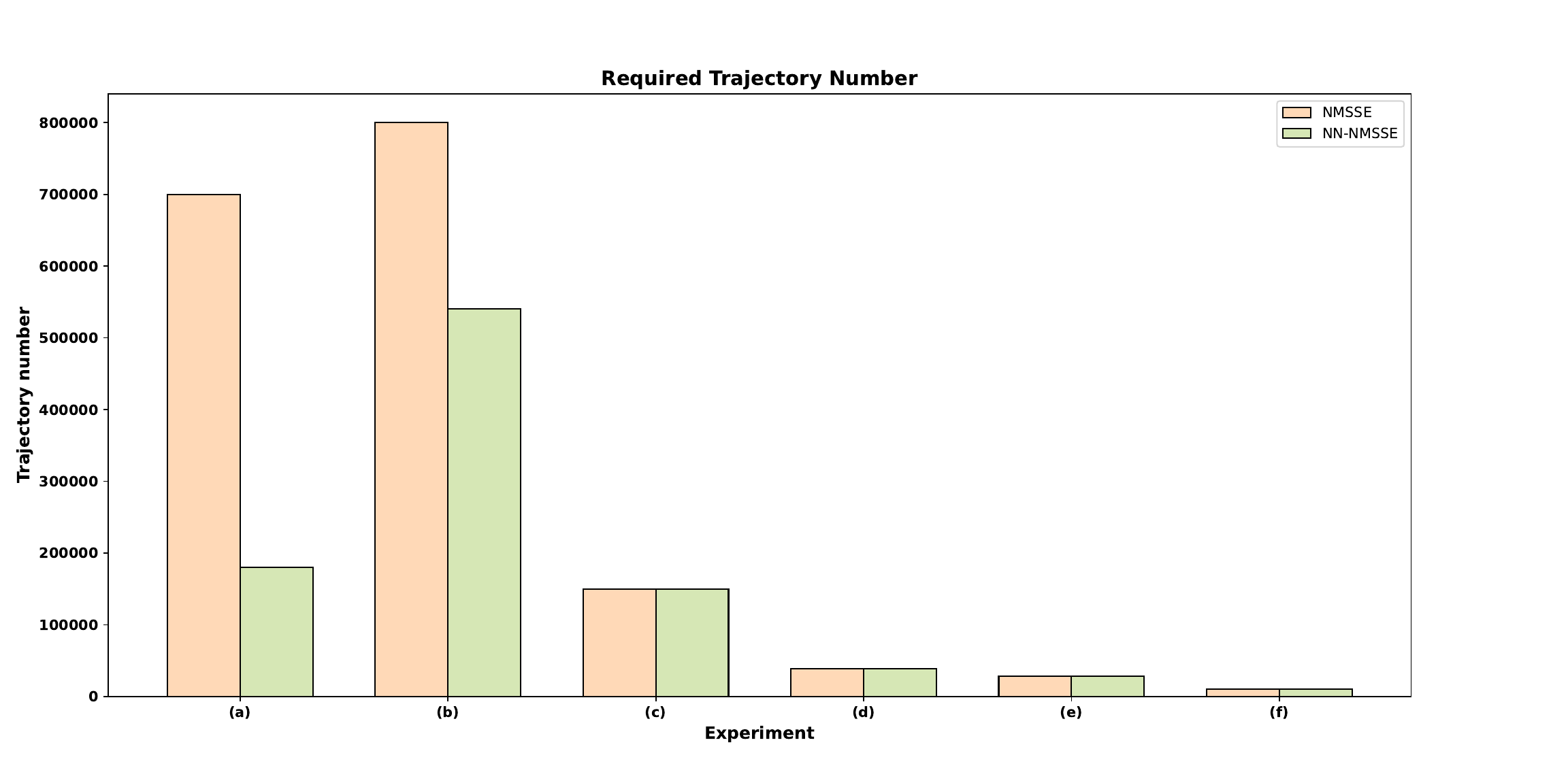}
    \caption{To achieve the same accuracy as the NN-NMSSE method, the number of trajectories for the NMSSE method must be increased significantly. The figure illustrates the required trajectory counts.}
    \label{fig:cost}
\end{figure}

\begin{figure}
    \includegraphics[scale=0.45]{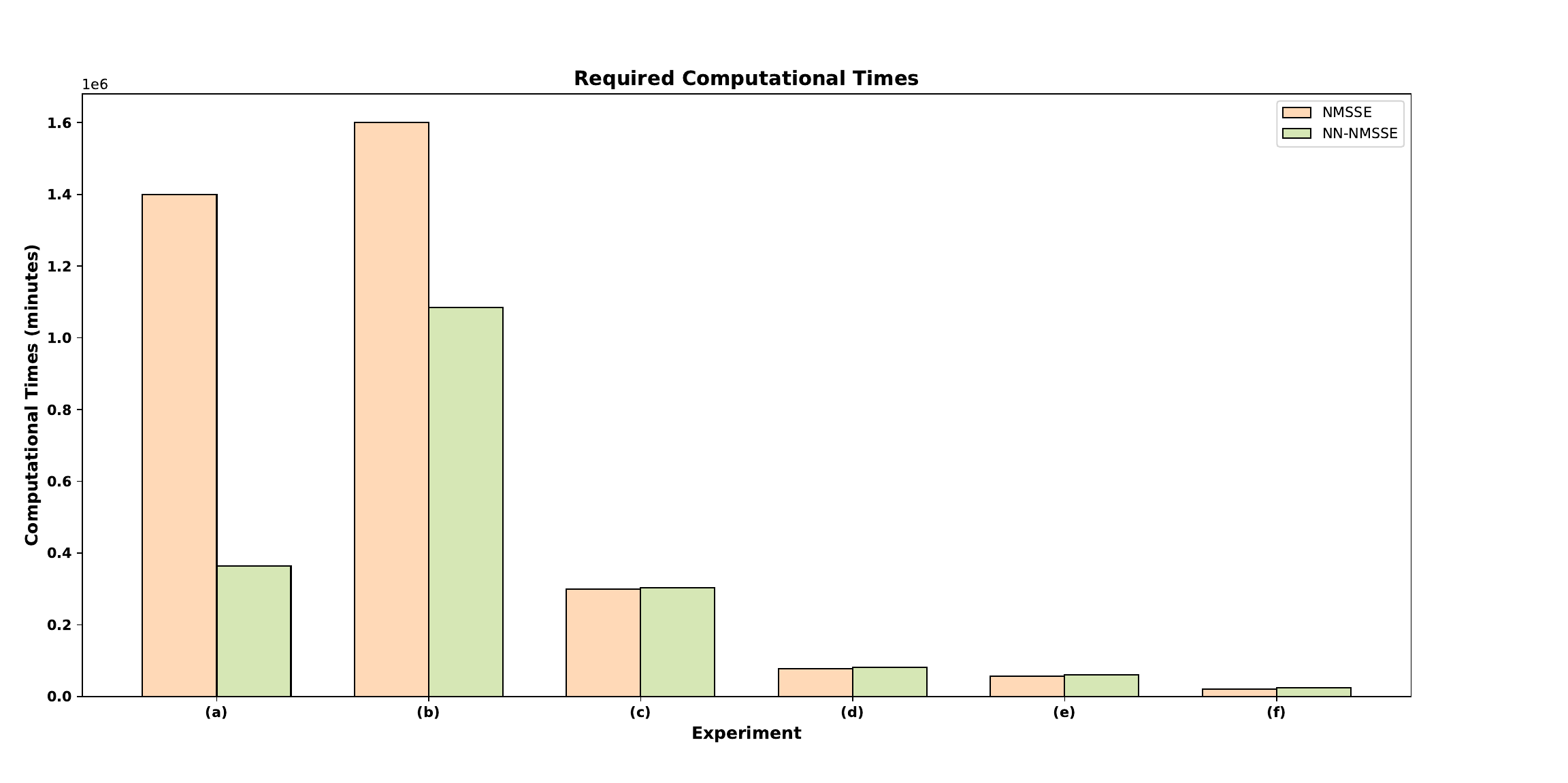}
    \caption{To achieve the same accuracy as the NN-NMSSE method, the number of trajectories for the NMSSE method must be increased significantly. The figure illustrates the required computational times. All time values are reported in minutes.}
    \label{fig:cost-time}
\end{figure}

To conduct a more detailed cost comparison, we focus on experiment (a), with results for experiments (b) to (f) summarized in Figure~\ref{fig:cost-time}.

For experiment (a), simulating a single stochastic trajectory with the NMSSE method requires 2 minutes. Achieving the desired accuracy without incorporating neural networks necessitates 700,000 trajectories, resulting in a total computation time of 1,400,000 minutes. In contrast, the NN-NMSSE approach reduces the number of required trajectories to 180,000. Training a single deep neural network model takes approximately 15 minutes. During a grid search, 25 models with varying neuron counts and input vector lengths were trained, consuming 375 minutes. To assess convergence, 10 groups of stochastic trajectories were generated, resulting in a total training time of 3,750 minutes. Long-term propagation predictions required an additional 30 minutes per group, equating to 300 minutes for all 10 groups. In total, the NN-NMSSE scheme required 360,000 minutes for stochastic trajectory simulations, and 4,050 minutes for neural network training and predicting, resulting in a cumulative computation time of 364,050 minutes.

\begin{figure}
    \centering
    \includegraphics[scale=0.45]{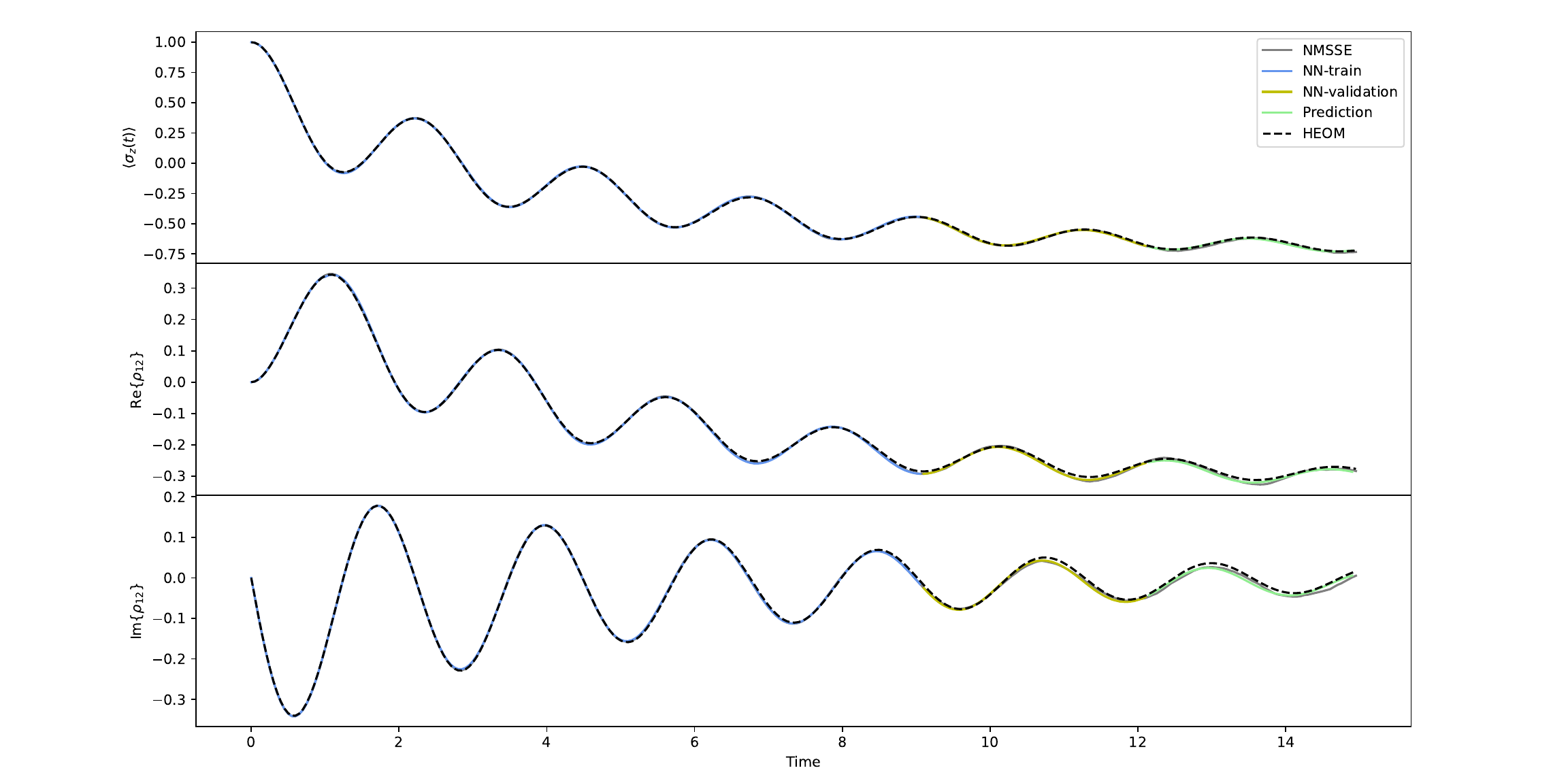}
    \caption{The figure illustrates the NN-NMSSE predictions for shorter evolution times.}
    \label{fig:short}
\end{figure}

The NMSSE method requires 1,400,000 minutes to achieve the desired accuracy, whereas the NN-NMSSE method reduces this time to 364,050 minutes—nearly a fourfold decrease. This demonstrates the significant advantage of NN-NMSSE in open quantum simulations, as it substantially reduces both the number of required trajectories and the overall computation time.

The computational costs for other experiments are summarized in Figure~\ref{fig:cost-time}. As shown, the required number of trajectories and computation times are significantly reduced for experiments (a) and (b), which correspond to low-temperature and non-adiabatic scenarios.

This approach is also effective for shorter evolution times. For example, the NN-NMSSE method was tested over the time interval $t = 0.0  $ to $ t = 15.0$, for experiment (a), with predictions closely matching the HEOM benchmark, as shown in Figure~\ref{fig:short}. These results confirm that the NN-NMSSE method performs reliably across varying propagation times.

\subsection{Excitation Energy Transfer in FMO Complex
}

Building on the successful results demonstrated for the spin-Boson model, the NN-NMSSE method was further validated on the more complex problem of excitation energy transfer (EET).

The system Hamiltonian for the molecular aggregate model with 
$N$ sites is expressed in the form of a Frenkel exciton model:
$H_{s}=\sum_{n=1}^{N}E_{n}|n\rangle\langle n|+\sum_{n\neq m}V_{nm}|n\rangle\langle m|$, where $|n\rangle$ represents the state in which only the $n^{\text{th}}$ site is excited with energy $E_n$. The term $V_{nm}$ denotes the coupling strength between the $n^{\text{th}}$ and $m^{\text{th}}$
sites. Assuming each site is coupled to an independent bath \cite{couplesite}, the system-bath coupling operator for each site is given by $\hat{x}_{n}=|n\rangle\langle n|$. The phonon properties are described by the Debye-Drude spectral density:
$J(\omega)=\frac{2\lambda\omega\gamma}{\omega^{2}+\gamma^{2}}$. The reorganization energy $\lambda$ is set to $\lambda=35\mathrm{cm}^{-1}$. The characteristic frequency $\gamma$ is determined by the inverse of the characteristic time scale, $\tau_{c}$, with $\gamma=\tau_{c}^{-1}=\frac{1}{50\mathrm{~fs}}$.

\begin{figure}[h]
    \centering
    \includegraphics[scale=0.45]{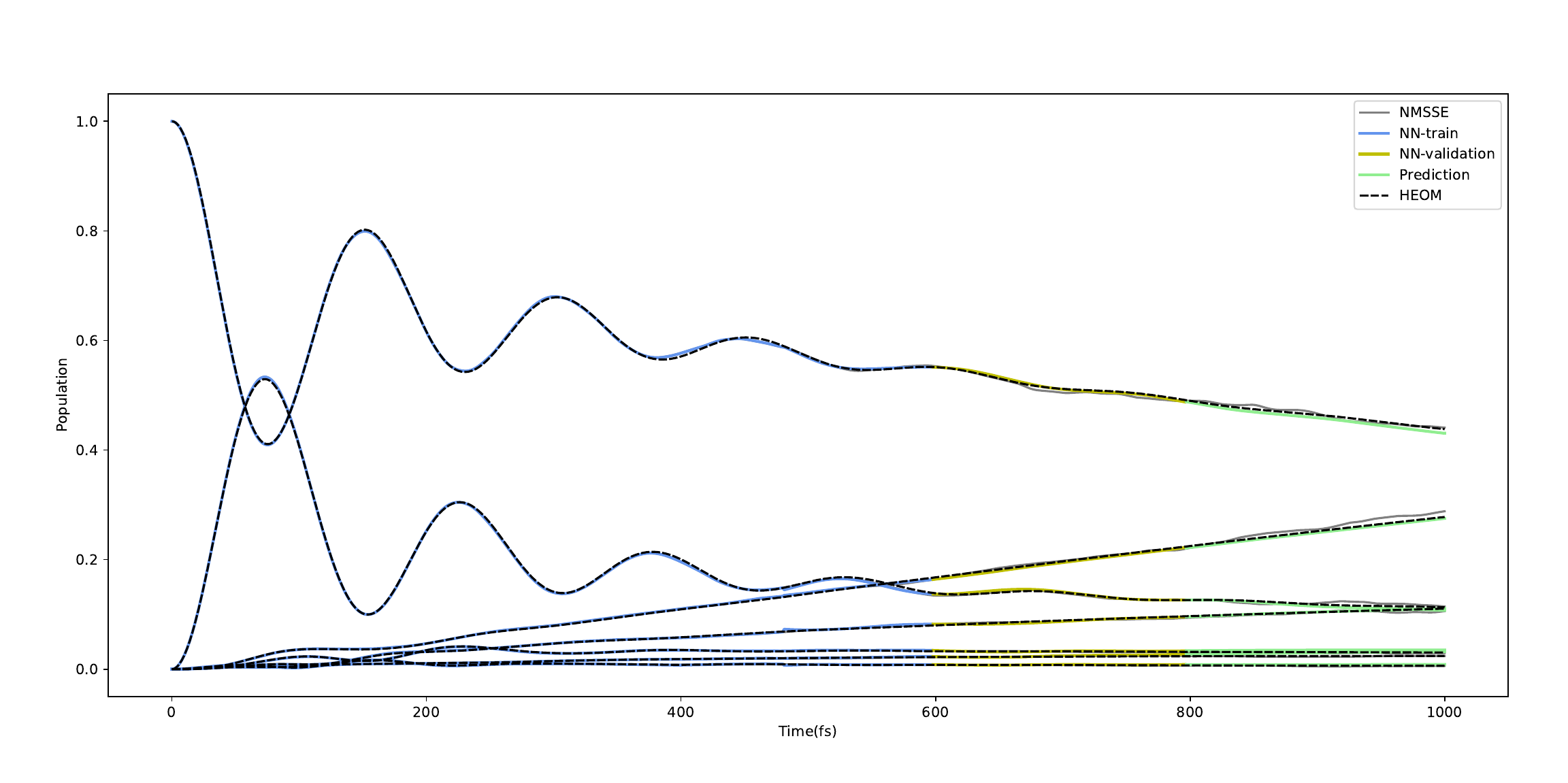}
    \caption{The figure illustrates excitation energy transfer in the FMO complex as predicted by the NN-NMSSE method at 77K. These results are compared with NMSSE simulations and benchmark data from HEOM. The simulations capture the dynamics over a timespan of 0 to 1000 femtoseconds.}
    \label{fig:fmo_low}
\end{figure}

Figure~\ref{fig:fmo_low} presents the propagation of excitation energy transfer predicted by the NN-NMSSE method, compared with the HEOM benchmark at a temperature of $T=77$K. As observed, stochastic oscillations arise in NMSSE simulations. By employing the NN-NMSSE approach, these oscillations are effectively mitigated, yielding stable and reliable results.

\begin{figure}
    \centering
    \includegraphics[scale=0.45]{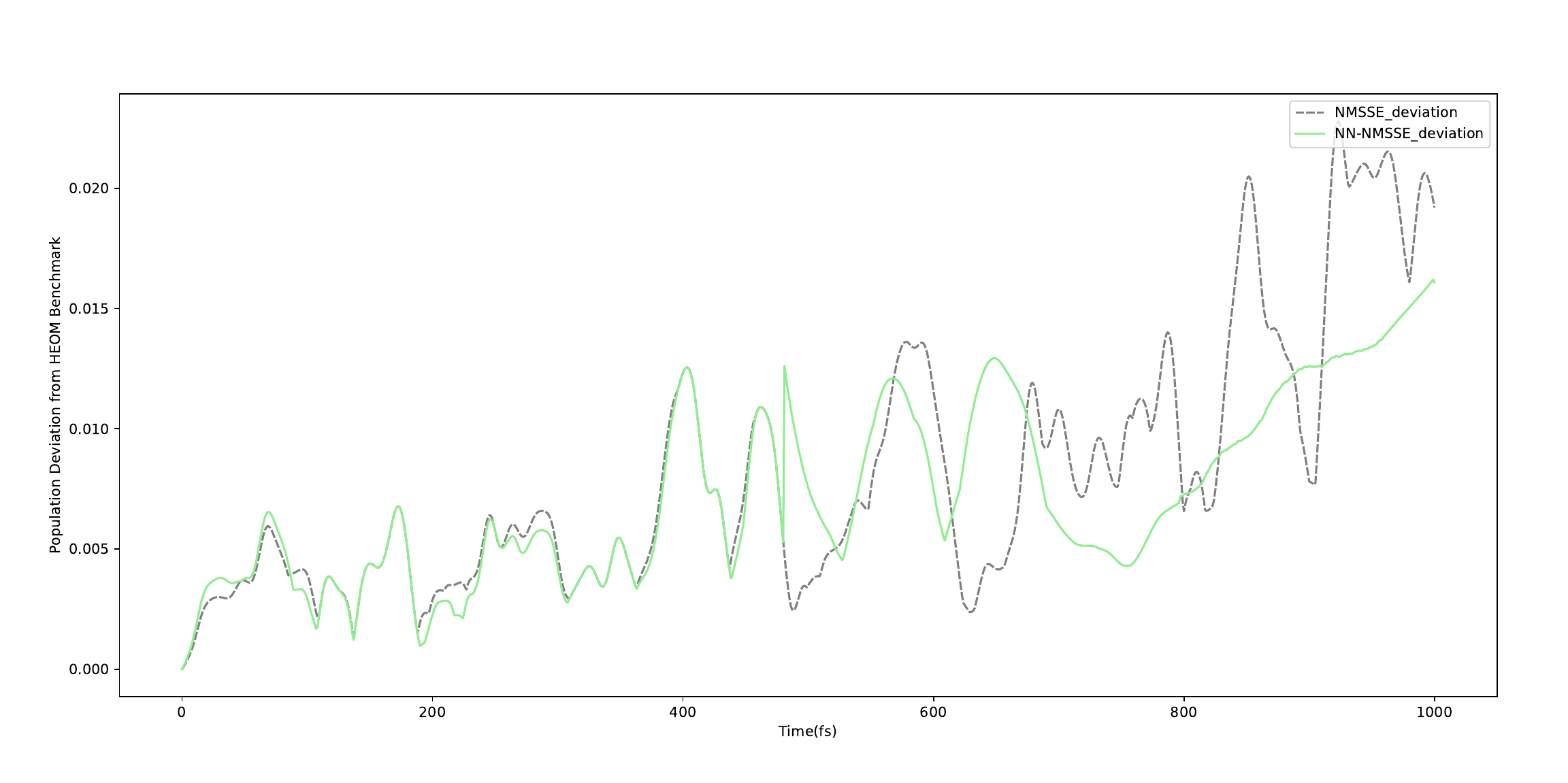}
    \caption{We compares the performance of NMSSE and NN-NMSSE method with 30,000 trajectories. The figure illustrates the deviations between the two methods and the exact results from the HEOM benchmark. The deviations for the NN-NMSSE approach are represented by green lines, while those for the NMSSE simulation method are shown in gray.}
    \label{fig:fmo_low_de}
\end{figure}

We calculated the deviations from the HEOM benchmark, comparing the simulations of NMSSE with the predictions of NN-NMSSE. The results are illustrated in the Figure~\ref{fig:fmo_low_de}. We simulated 30,000 trajectories using cNMSSEs. As shown in the Figure~\ref{fig:fmo_low_de}, NN-NMSSE can reduce $25\%$ of the deviation utilizing the same number of trajectories. In further testing, achieving the same low deviation only with NMSSE required an additional 8,000 trajectories. Table~\ref{tab:fmo_cost} compares the computational cost of the NMSSE method without neural network assistance to that of the NN-NMSSE method.

\begin{figure}
    \includegraphics[scale=0.45]{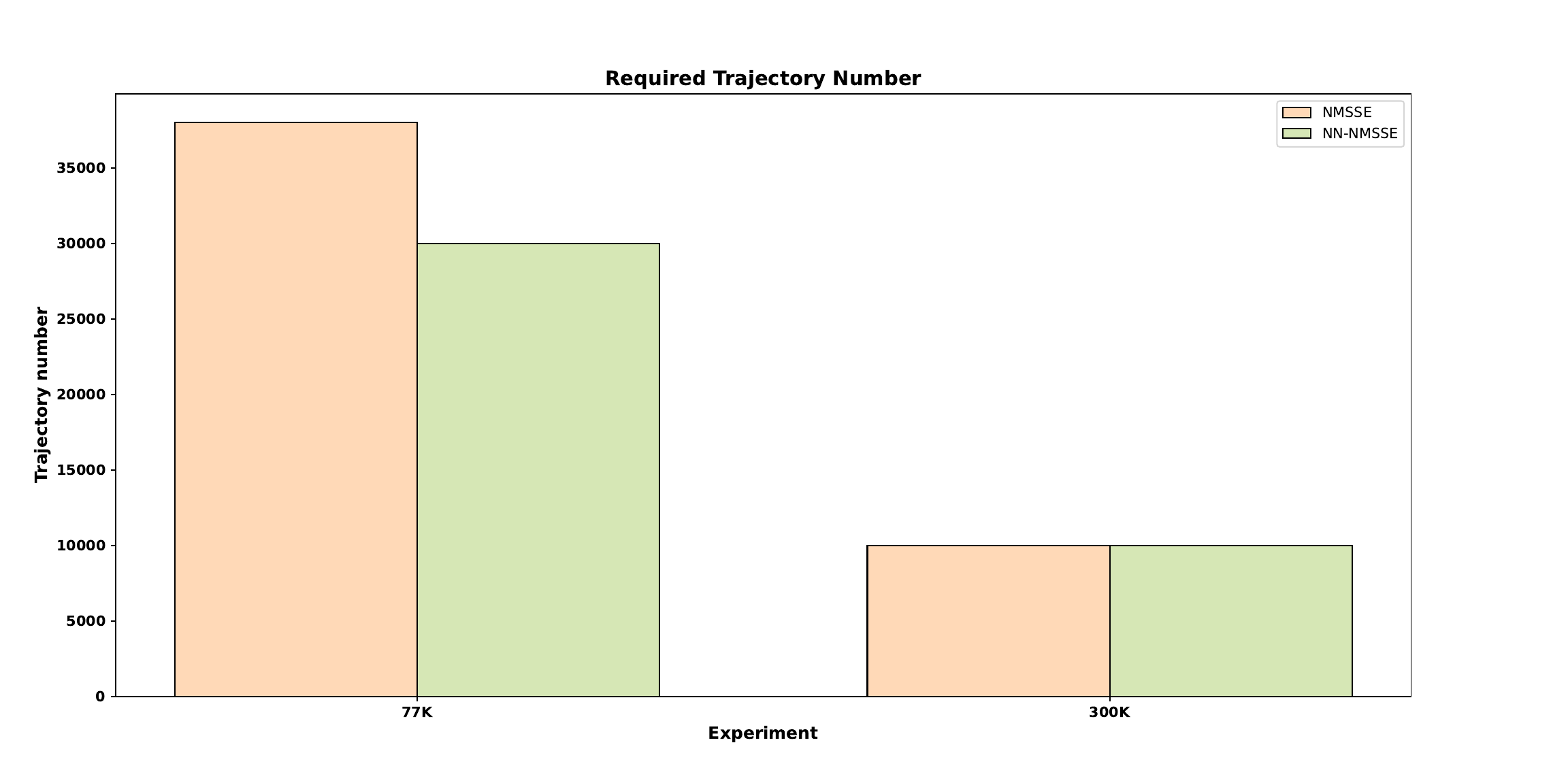}
    \caption{To achieve the same accuracy as the NN-NMSSE method, the number of trajectories for the NMSSE method must be increased significantly. The figure illustrates the required trajectory counts.}
    \label{fig:cost}
\end{figure}

\begin{figure}
    \includegraphics[scale=0.45]{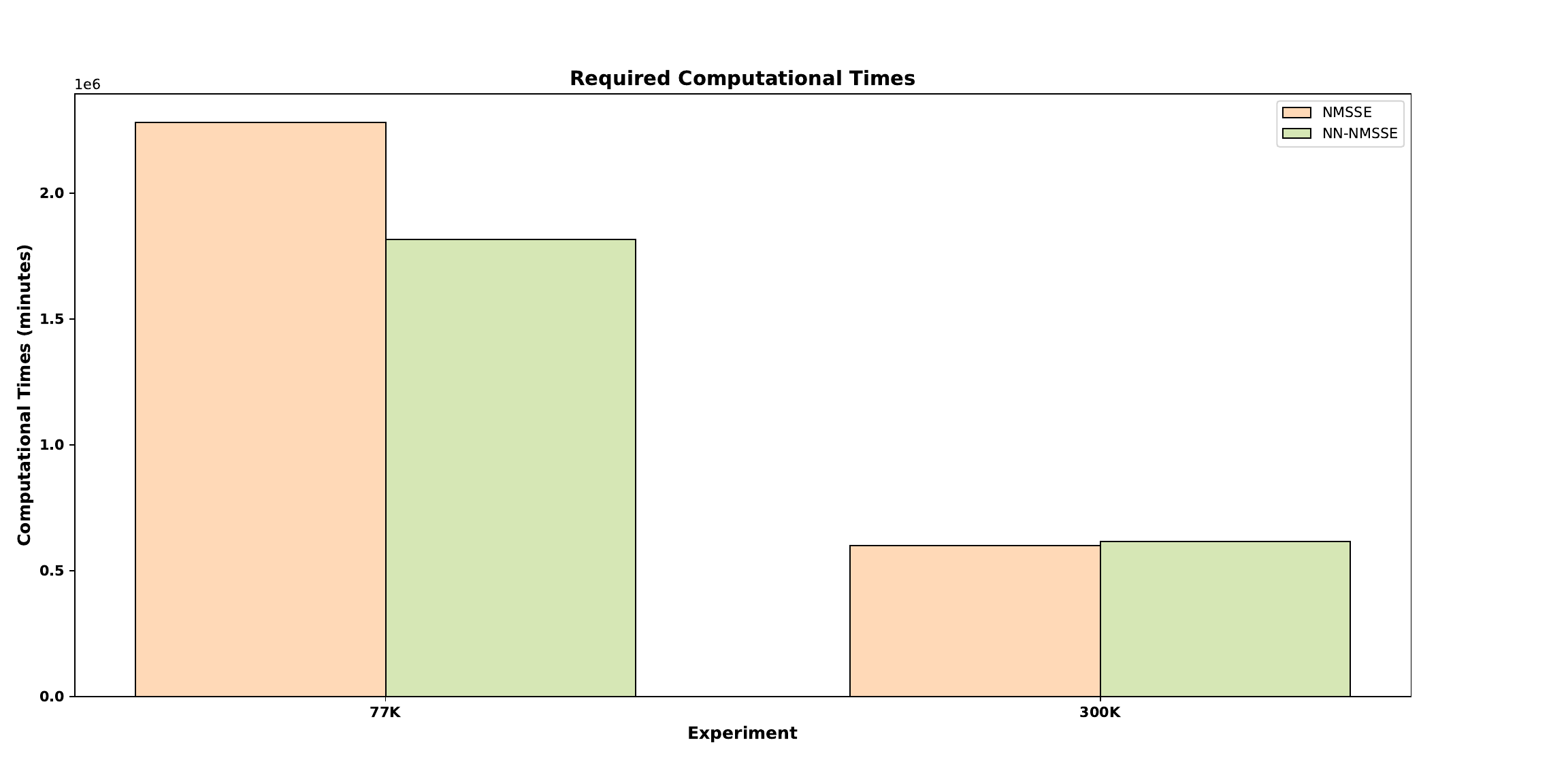}
    \caption{To achieve the same accuracy as the NN-NMSSE method, the number of trajectories for the NMSSE method must be increased significantly. The figure illustrates the required computational times. All time values are reported in minutes.}
    \label{fig:cost-time}
\end{figure}

At temperature $77$K, simulating a single stochastic trajectory with the NMSSE method requires 60 minutes performed on our workstation. Achieving the desired accuracy without incorporating neural networks necessitates 38,000 trajectories, resulting in a total computation time of 2,280,000 minutes. In contrast, the NN-NMSSE approach reduces the number of required trajectories to 30,000. Training a single deep neural network model takes approximately 60 minutes. During a grid search, 25 models with varying neuron counts and input vector lengths were trained, consuming 1,500 minutes. To assess convergence, 10 groups of stochastic trajectories were generated, resulting in a total training time of 15,000 minutes. It required an additional 45 minutes to predict propagations using the optimal neural network per group, equating to 450 minutes for all 10 groups. In total, the NN-NMSSE scheme required 1,800,000 minutes for stochastic trajectory simulations, and 15,450 minutes for neural network training and predicting, resulting in a cumulative computation time of 1,815,450 minutes.

The NMSSE method requires 2,280,000 minutes to achieve the desired accuracy, whereas the NN-NMSSE method needs only 1,815,450 minutes. This demonstrates the significant advantage of NN-NMSSE in open quantum simulations, as it substantially reduces both the number of required trajectories and the overall computation time.

The computational cost for high temperature ($300$K) experiment is summarized in Table~\ref{tab:fmo_cost}. For high-temperature scenarios, NN-NMSSE can not improve the efficiency. However, the required number of trajectories are much smaller compared to low-temperature scenarios. So there is no need to use NN-NMSSE method.

\begin{figure}
    \centering
    \includegraphics[scale=0.45]{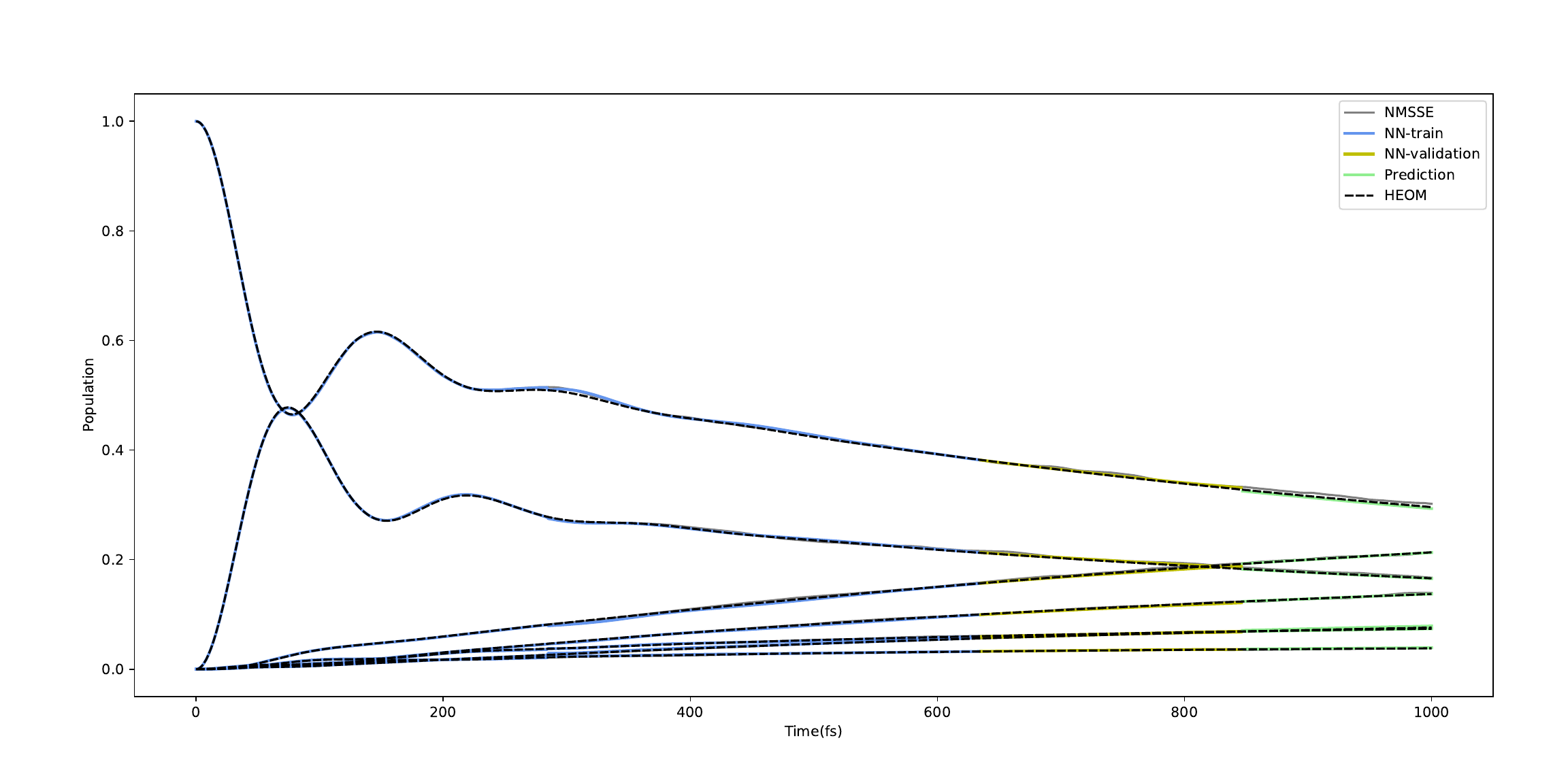}
    \caption{The figure illustrates excitation energy transfer in the FMO complex as predicted by the NN-NMSSE method at 300K. These results are compared with NMSSE simulations and benchmark data from HEOM. The simulations capture the dynamics over a timespan of 0 to 1000 femtoseconds.}
    \label{fig:fmo_high}
\end{figure}

\begin{figure}
    \includegraphics[scale=0.45]{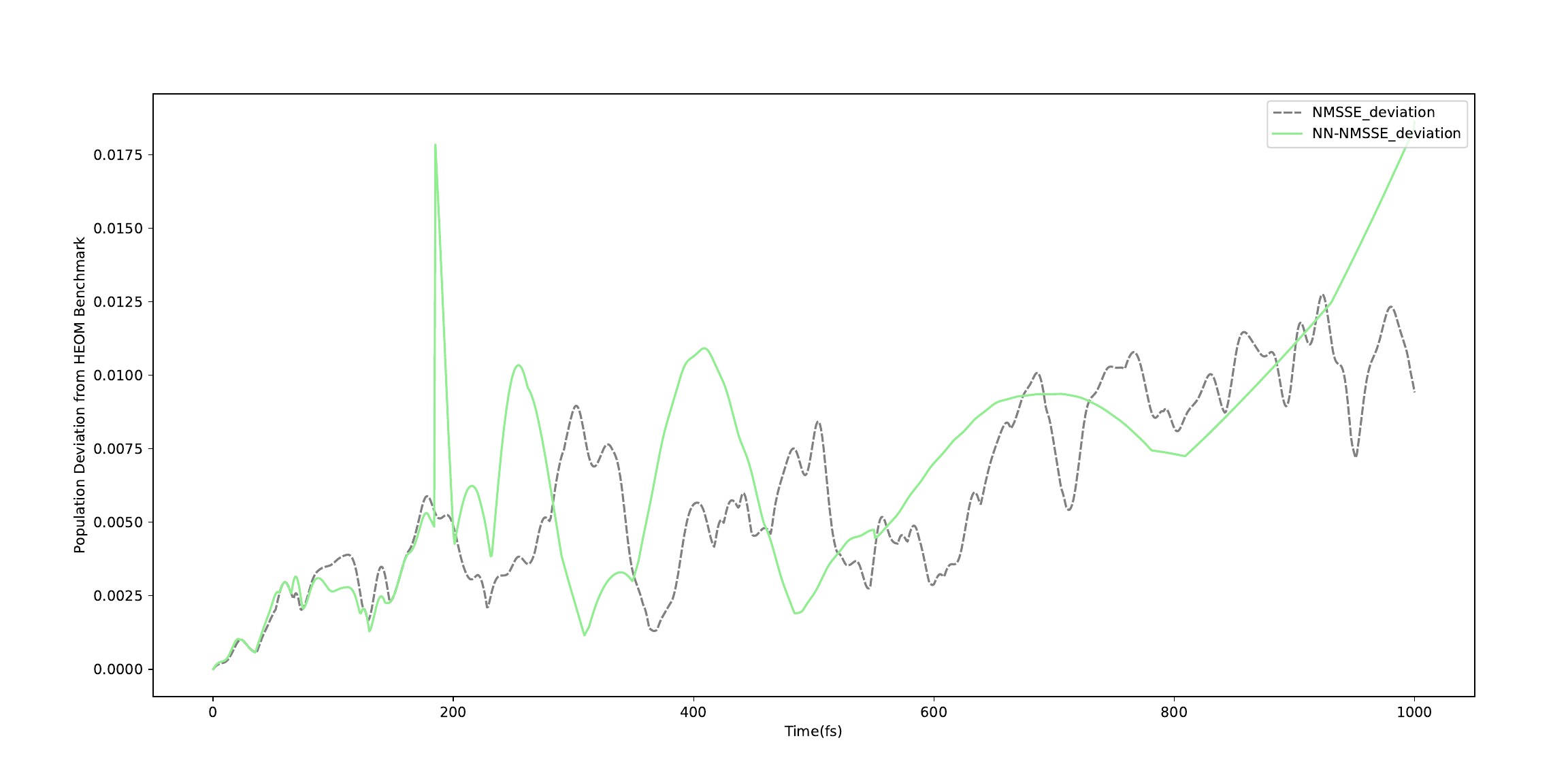}
    \caption{We compares the performance of NMSSE and NN-NMSSE method with 10,000 stochastic trajectories at 300K. The figure illustrates the population deviations between two methods and the exact results from the HEOM benchmark.}
    \label{fig:fmo_high_de}
\end{figure}

These numerical experiments demonstrate that the NN-NMSSE method is highly effective across different models under various parameter conditions. Notably, at low temperatures, NN-NMSSE significantly enhances the efficiency of the NMSSE method by reducing the number of required stochastic trajectories and lowering computational costs. Before employing the NN-NMSSE method, it is useful to evaluate the convergence behavior of the stochastic trajectories. If the standard error grows over time with an accelerating rate, it indicates that the dynamical model experiences severe stochastic oscillations in numerical simulations, making convergence difficult. In such cases, the NN-NMSSE method is highly advantageous, significantly reducing the number of required trajectories and the associated computational cost.

On the other hand, if the standard error does not increase rapidly over time, the model is unlikely to encounter major convergence issues, and the use of the NN-NMSSE method may not be necessary.

\section{Conclusion}

The primary focus of this paper is addressing the challenge of stochastic oscillations and the resulting convergence difficulties in simulating the dynamics of open quantum systems using stochastic quantum state diffusion type methods.

By leveraging the compatibility between stochastic quantum state diffusion methods and time-series neural networks, the NN-NMSSE method capitalizes on the intrinsic statistical convergence of stochastic trajectories. This synergy allows neural network predictions to be anchored to statistically reliable references, enhancing the credibility of the approach. This reliability sets it apart from other neural network-based dynamic methods.

Furthermore, the NN-NMSSE method resolves the issue of stochastic oscillation convergence. It significantly reduces the number of stochastic trajectories and computational cost required by traditional methods, thereby enabling more efficient simulations of larger systems and the exploration of steady-state properties in open quantum systems.

The NN-NMSSE method demonstrates exceptional performance under low-temperature and non-adiabatic conditions, substantially reducing computational costs and addressing stochastic oscillation issues. However, in regions where convergence is not problematic, NN-NMSSE does not outperform the original NMSSE method. As analyzed in the Numerical Results section, the advantage of NN-NMSSE becomes evident only when the standard error of the stochastic trajectories increases significantly with time evolution, indicating a convergence challenge. Thus, NN-NMSSE is a specialized approach designed specifically to tackle the issue of stochastic oscillations.

\begin{acknowledgement}

The authors thank Qiang Shi for sharing his code in HEOM; The authors thank Qiang Shi for sharing his code in HEOM; We thank Jiajun Ren for the helpful discussions about Renormalizer.

\end{acknowledgement}

\begin{suppinfo}

Several relevant information: the derivation of non-Markovian stochastic Schrödinger equations with complex modes(cNMSSEs); the detailed structure of neural networks with varying neuron counts and input vector lengths for experiments mentioned in the Numerical Result section; the loss function and the optimizer for training neural networks; more test results on different quantum systems.

\end{suppinfo}

\bibliography{achemso-demo}

\end{document}